\documentclass[jap,aip,twocolumn,10pt,groupedaddress]{revtex4-1}
\usepackage[table,x11names]{xcolor}
\usepackage{url}		% Allow URLs to be display
\usepackage{graphicx}		% The most modern graphics package
\usepackage{dcolumn}		% Decimal alignment package
\usepackage{graphicx}		% Graphics package
\usepackage{dcolumn}		% Decimal alignment package
\usepackage{amsmath}
\usepackage{amssymb}
\usepackage{latexsym}		% Symbols
\usepackage[absolute,overlay]{textpos}
\usepackage[T1]{fontenc}
\usepackage{avant}
\usepackage{hyperref}
\usepackage{makecell}
\usepackage{textcomp}
\usepackage{xfrac}
\usepackage{xcolor} %colors,yay

\makeatletter

\newcommand{\Rmnum}[1]{\expandafter\@slowromancap\romannumeral #1@}
\makeatother

\makeatletter
\g@addto@macro\bfseries{\boldmath}
\makeatother

\makeatletter

    \def\CT@@do@color{%
      \global\let\CT@do@color\relax
            \@tempdima\wd\z@
            \advance\@tempdima\@tempdimb
            \advance\@tempdima\@tempdimc
    \advance\@tempdimb\tabcolsep
    \advance\@tempdimc\tabcolsep
    \advance\@tempdima2\tabcolsep
            \kern-\@tempdimb
            \leaders\vrule
                    \hskip\@tempdima\@plus  1fill
            \kern-\@tempdimc
            \hskip-\wd\z@ \@plus -1fill }
\makeatother

\begin{document}

%\title{Demonstration of a Cargo Scanning System Prototype Using Multiple  Monoenergetic Gamma Radiography (MMGR) for Effective Atomic Number Identification}
\title{Experimental Demonstration of Multiple Monoenergetic Gamma Radiography for Effective Atomic Number Identification in Cargo Inspection}
\author{Brian S. Henderson}
\email[E-Mail: ]{bhender1@mit.edu}
\author{Hin Y. Lee}
\author{Thomas D. MacDonald}
\author{Roberts G. Nelson}
\author{Areg Danagoulian}
\affiliation{Department of Nuclear Science and Engineering, Massachusetts Institute of Technology, Cambridge, Massachusetts 02139}
\date{\today}

\begin{abstract}

%Special nuclear material, cargo security, etc.

%Focus on fact that results show precise determination of $Z$ and areal density, even at high-$Z$, could plausibly separate benign
%lead from SNM.

The smuggling of special nuclear materials (SNM) through international borders could enable % Edit: abbreviation up here now
nuclear terrorism and constitutes a significant threat to global security. This 
paper presents the experimental demonstration of a novel radiographic technique for 
%scanning cargo to  % changed this to commercial cargoes
quantitatively reconstructing the density and type of material present in commercial cargo containers, as a means of detecting such threats.
%This technique utilizes monoenergetic photons from nuclear reactions, specifically the 4.4 and 15.1~MeV % Edit: ~
%photons from the $^{11}$B(d,n$\gamma$)$^{12}$C reaction, rather than a continuous spectrum
%bremsstrahlung photon source as is typically used for cargo radiography, to achieve strong
%resolving power to differentiate materials.
Unlike traditional techniques which use sources of bremsstrahlung photons with a continuous distribution of energies, 
multiple monoenergetic gamma radiography (MMGR) utilizes monoenergetic photons from nuclear reactions, specifically the 4.4 and 15.1~MeV % Edit: ~
photons from the $^{11}$B(d,n$\gamma$)$^{12}$C reaction.
By exploiting the $Z$-dependence of the photon interaction cross sections at these two % Edit: Math Z for consistency, did a search fix for this throughout
specific energies it is possible to simultaneously determine the areal density and the effective
atomic number as a function of location for a 2D projection of a scanned object. %cargo->object
The additional information gleaned from using and
detecting photons of specific energies for radiography substantially increases the resolving power between
different materials.
This paper presents results from the imaging of mock cargo materials ranging 
from $Z\approx5\text{--}92$, demonstrating accurate
reconstruction of the effective atomic number and areal density of the materials 
over the full range.  In particular, the system
is capable of distinguishing pure materials with $Z\gtrsim70$, such as lead and % Edit: Proper emdash
uranium --- a critical requirement of a system designed to detect SNM.  This 
methodology could be used to screen commercial cargoes with high material specificity, to distinguish most benign materials from SNM, such as % Edit: Deleted ``Thus''
uranium and plutonium.

% As of 1-24, abstract is 228 words (JAP recommends max of ~250, no hard limit)

\end{abstract}

\maketitle{}

\section{Introduction}

The field of nuclear security addresses the danger of the nuclear weapons, 
including proliferation of weapon technology, safeguards of 
fissile materials, and the risk of nuclear terrorism.  The latter topic encompasses cargo security, which specifically focuses 
on preventing the smuggling of nuclear materials and fully assembled nuclear 
devices through ports of entry and other pathways. %<-- added that. 
Estimates of the immediate economic costs alone
of a nuclear explosion in a major port exceed \$1 trillion, \textit{before} accounting for the substantial human costs \cite{randecon,abtes}.
Given the relative anonymity of cargo shipping and its resulting vulnerability to smuggling, the
lack of systems to efficiently and reliably %<-- added
deter nuclear smuggling %<- in the flow of commerce, from in cargo
 remains a relevant security threat.   This paper details the demonstration
of a new radiography technique for quantitatively identifying materials in cargo that is capable of distinguishing different
materials with high atomic number. Specifically, the technique is capable of separating benign high-$Z$ materials
such as lead and tungsten from special nuclear materials (SNM).

\subsection{Detecting Nuclear Material in Cargo}

Approximately 40000--57000 maritime shipping containers enter the United States every % Edits: Wording, Estimated->Approximately; proper endash, removed ISO
day~\cite{kouzes}.  This fast throughput rate and the fact
that many containers are densely packed to weights of up to 20 metric tons
make cargo containers particularly vulnerable to the smuggling of nuclear
materials or weapons.  A system designed to detect nuclear smuggling must
simultaneously achieve the following: scan cargo at $\lesssim$1 minute per container,
produce a low rate of false positives, and provide a clear indicator of the presence of nuclear materials
in diverse cargo configurations (i.e., a low rate of false negatives). %<- reworded this sentence  
Additionally, port operations restrict the footprint of scanning systems, as well as the % CHANGE
permissible radiation dose to the cargo and surrounding area \cite{cbo,wco}.

Cargo screening technologies can be classified into three categories:  
passive interrogation,  active interrogation, and  radiography.  Passive 
interrogation involves the detection of the natural radioactivity of various 
particles --- neutrons and photons in particular --- from fissile materials.  In 
this context the materials of interest are primarily weapons grade uranium 
(WGU), which consists primarily of $^{235}$U, and weapons grade plutonium (WGPu).  % Edit: Missing space
The later primarily consists of $^{239}$Pu, but its other isotopes 
($^{240}$Pu in particular) play a key role in its passive signature.  

Passive detection systems offer simplicity and relatively low cost, and such systems have
been deployed widely in the United States and elsewhere.
These systems primarily consist of portals, which use various scintillators
to detect photons in combination with $^3$He neutron detectors to uncover
SNM by their radioactive emissions from nuclear decay and spontaneous fission.
The addition of shielding around smuggled material, however, circumvents
passive detection. The passive signal from WGU is very weak and easily
shielded, while even an assembled plutonium device (with its strong spontaneous
fission neutron signature) may be shielded with combinations of low- and high-$Z$
material to block both neutron and photon signals.

% from spontaneous fission, 
% beta decay of the fission products, or photons from the excited states of alpha 
% decay products. The spontaneous fission neutrons are detected either by $^3$He 
% tubes or some other neutron detector.  They also have significant weaknesses.  
% WGU for example has very low spontaneous fission rate, and its most prominent 
% photon line is 186 keV, which can be easily shielded by most cargoes.

% WGPu, depending on $^{240}$Pu concentration, emits neutrons at the approximate rate of 
%  70000 kg$^{-1}$s$^{-1}$.  While significant, this emission signature can be 
% readily shielded by moderate amounts of borated polyethylene, followed by modest 
% amounts of lead or steel to shield the photons from (n,$\gamma$) neutron capture 
% on boron \comment{(might be useful to list the actual dimensions, from Brian's 
% simulations)}.

The limitations of passive detection techniques necessitate alternative 
approaches.  Active interrogation systems expose the cargo to a beam of one
or more types of particles (such as photons, neutrons, or muons) to trigger
secondary processes unique to fissionable and fissile materials, producing
signals which are strong enough to overcome shielding attempts by a competent smuggler.
Examples of such systems include prompt neutrons from photofission (PNPF)~\cite{pnpf-short}, EZ3D~\cite{ez3d_patent}, 
and nuclear resonance fluorescence~\cite{NRF_bertozzi}.  Furthermore, other 
groups related to this research effort have advanced the detection of delayed 
neutrons from induced fission as a way of identifying fissile 
materials~\cite{mayer}.  While such techniques have promise due to their specificity for SNM detection,
no system has been sufficiently developed for deployment at this time.  For a high level discussion
of several active detection methodologies see Runkle, \textit{et al}.~\cite{runkle2012rattling}.    % Edit: format

\subsection{Radiography for SNM Detection}

While searching for shielded fissile and fissionable materials via active 
methods is promising, shielding scenarios which completely block the signal are 
nevertheless possible.  Radiographic imaging of cargo provides a means of
detecting such scenarios.
A variety of radiographic techniques have been proposed in the past, 
including using medium energy ($\sim$GeV) protons~\cite{king1999800}, muons\cite{schultz2004,morris2008,borozdin2003}, neutrons\cite{jill,sowerby,cutmore}, as well as 
$\sim$keV photons (X-rays) and $\sim$MeV photons (gamma rays) \cite{chen2007,gilbert_spectral_analysis}.  Additionally, radiographic
imaging of cargo for SNM overlaps well with other goals of cargo inspection
(such as detection of non-nuclear smuggling), adding value to the technique.

This work builds upon prior studies of using 4.4 and 
15.1~MeV monochromatic photons from the $^{11}$B$(d,n\gamma)^{12}$C reaction to 
radiograph objects and differentiate between their material 
types\cite{oday2015,buck}.  
A parallel effort by other groups, using Cherenkov detectors, have used the same 
reaction to pursue a similar goal \cite{rose}.  The prior work however did not achieve a precise
determination of the atomic number or areal density of the scanned objects.  This work demonstrates the 
ability to infer the effective atomic number ($Z$) and areal density ($\rho_A$) of a given 
spatial pixel across a cargo sample, providing essential information for the identification
of materials present in the cargo.  This reconstruction is shown to be accurate enough to 
distinguish between uranium and lead, a critical result for SNM detection in that it
permits distinguishing nuclear threats from benign materials.  With this capability,
the system is robust against false alarm scenarios in which benign high-$Z$ materials (e.g., lead,
tungsten, precious metals) appear similar to SNM and thus require further inspection.

In its simplest form radiography combines measurements the transmitted photon 
flux $\phi$ for a given material sample, knowledge of the incident flux $\phi_0$,
and an assumption of the mass attenuation coefficient $\mu$ of the material to infer an approximate areal density 
$\rho_A$:
\begin{equation}
\mu \rho_A = \ln{(\phi_0/\phi)} ~. \label{eq:dmu}
\end{equation}
This calculation can be performed for every pixel in a radiographic scan to image
the sample. By assuming that $\mu$ does not vary through the scan plane, a {\it 
relative} value of $\rho_A$ can be reconstructed. It should be noted that $\mu$ depends on the
elemental composition of the material, and thus is a function of effective atomic number $Z$.  As such, a measurement such as this cannot allow a simultaneous determination 
of effective atomic number $Z$ and areal density $\rho_A$, a requirement for distinguishing SNM from benign cargo. 

%The radiographic detection of smuggled SNM among benign cargo requires determination of both effective $Z$ and $\rho_A$. 
This goal can be achieved by using the energy dependence of $\mu$, and 
performing multiple measurements at various energies.  The main processes which 
contribute to photon attenuation at 4.4 and 15.1 MeV are Compton scattering and 
pair production.  The mass attenuation coefficient can be approximated as $\mu = 
\mu_{c} + \mu_{pp}$, where $\mu_{c}$ and $\mu_{pp}$ are the coefficients for 
Compton scattering and pair production, respectively.  Each of these coefficients
depends on $Z$ and the incident photon energy $E$ in different ways.  Specifically,
\begin{align*} 										% Edit: Allowed these equations to be numbered
\mu_c = Z N_A \sigma_{c}(E,Z) / A \\
\mu_{pp} = N_A \sigma_{pp}(E,Z) / A,
\end{align*}
where $N_A$ is Avogadro's number, $A$ is the atomic weight of the material under inspection, and the $\sigma(E,Z)$ are the
cross sections of the relevant attenuation processes.  For photon energies satisfying $E \gg 511$ keV,
the cross sections may be approximated as $\sigma_c \propto 1/E$ and $\sigma_{pp} \propto Z^2 f(E_\gamma)$, where $f(E_\gamma)$ is a function of energy with
negligible dependence on atomic number \cite{Leo}. Using these as inputs to the mass 
attenuation coefficients to compute the transmission ratios (Equation~\ref{eq:dmu})
at two different energies ($E_0$ and $E_1$) results in % Edit: Equation written out
\begin{align}
R = \frac{\ln {(\phi(E_1)/\phi_0(E_1))}}{\ln{(\phi(E_0)/\phi_0(E_0) )}} &=  
\frac{ Z^2 f(E_1) N_A \cdot \mathrm{const_1}/A }{ Z N_A \cdot \mathrm{const_2}/AE_0}  \nonumber \\        % Edit: Fixed alignment
&= Z \cdot E_0f(E_1) C,  \label{eq:R}
\end{align}
where $C$ is a constant equal to the ratio $\mathrm{const_1/const_2}$.
This treatment assumes that at $E_0$ the mass attenuation is entirely dominated by 
Compton scattering, while at $E_1$ pair production dominates. Assuming these 
requirements are met, an experimental measurement of $R$ could be used to 
directly determine the atomic number $Z$ (and the total attenuation used to infer the areal density $\rho_A$). 
While this simple model requires broad approximations, Equation~\ref{eq:R} captures the essential mechanism
by which dual energy radiography may provide precise identification of the effective $Z$ of inspected materials.

\subsection{Monoenergetic Gamma Rays from Nuclear Reactions}

%How generated, advantages, other possible

%Advantages of monoenergetic photons

%Previous examples of use of the boron reaction

%Importance of operation at high rates, advantages of pulsed beam over CW

%Importance of dose, not only delivered directly to cargo but also in the surrounding area

%Make sure to get citations rounded-up for similar approaches (both in our ARI
%group and outside).

Dual energy radiography is by no means a new concept.  Current systems 
implement this technique by using bremsstrahlung beams with varying endpoints.  
Linear accelerator (linac) based bremsstrahlung dual energy systems typically vary the
electron beam energy between two fixed values (e.g., 6 and 9~MeV)\cite{chen2007,gilbert_spectral_analysis}. The transmitted signals are compared in a way that 
allows quantitative determination of the effective $Z$ of a given pixel in the cargo % Edit: Math Z
image\cite{tsinghua}.  Bremsstrahlung based systems, while capable of rapidly producing images with
excellent spatial resolution using commercially-produced equipment, have notable
disadvantages.  Most commercial linacs produce have a duty factor of $\sim$0.1\%, 
producing pulses of several {\textmu}s length at $\mathcal{O}(10^2\:\text{Hz})$.  The resulting large instantaneous flux prevents
measurement of the transmitted spectrum and only an integrated measurement of
the total deposited energy is possible.  This significantly reduces the information content of the signal,
increasing the number of photons (and thus radiation dose) required to reconstruct the material type.
Furthermore, most of the energy of the beam flux is at low energies ($\lesssim$1 MeV).  Photons at these
energies contribute to radiation dose, but provide little to no transmitted information due to the strong attenuation
at low energies.  For example, a Geant4\cite{geant} simulation 
shows that a system based on a 6 MeV electron beam would produce approximately 
90\% of the counts and 65\% of the radiation dose from photons $\leq$3 MeV. This  % Edit: \leq sign
translates to a low information-to-dose ratio.  Finally, a significant number of 
photons undergo scattering in the cargo but still reach the detectors,
which reduces the image contrast and dilutes the pixel-specific $Z$-dependent information content. % Edit: Math Z

Many of these factors can be overcome by replacing a linac-based system with one 
which uses nuclear reactions to produce monochromatic photons.  The technique
of using monoenergetic gamma rays at several energies, referred to as
multiple monoenergetic gamma radiography (MMGR), provides several advantages over
traditional bremsstrahlung radiography.
% Such sources can 
% be of much higher duty factor:  the DL-3 Radio Frequency Quadrupole (RFQ) accelerator % Edit: added abbreviation
% used in this study can achieve a 2\% duty, while the Ionetix ION-12$^\text{SC}$ % Edit: text SC superscript
% 12.5~MeV cyclotron has 100\% duty factor. \comment{Not sure the duty factors are important to the discussion right here}  This allows the use of spectroscopic 
% detector methodologies.
The knowledge of the photon energies and the measurement of transmitted spectral data allows the suppression of events
in the signal which have undergone scattering, thus leaving only the photons 
which have undergone direct line-of-sight transmission. This creates a clean transmitted
signal associated with each pixel, highly dependent on the effective $Z$ and areal density
of the intervening material.  This work utilizes the $^{11}$B(d,n$\gamma$)$^{12}$C 
reaction to produce 4.4 and 15.1~MeV photons, which arise from the short-lived 
excited levels of $^{12}$C in the final state of the reaction. The large spread in 
energy between the two gamma rays in the source spectrum provide strong leverage
for material identification.

\section{Experimental Methods}
\label{sec:exp}

\begin{figure}[htb]
\includegraphics[width=\columnwidth]{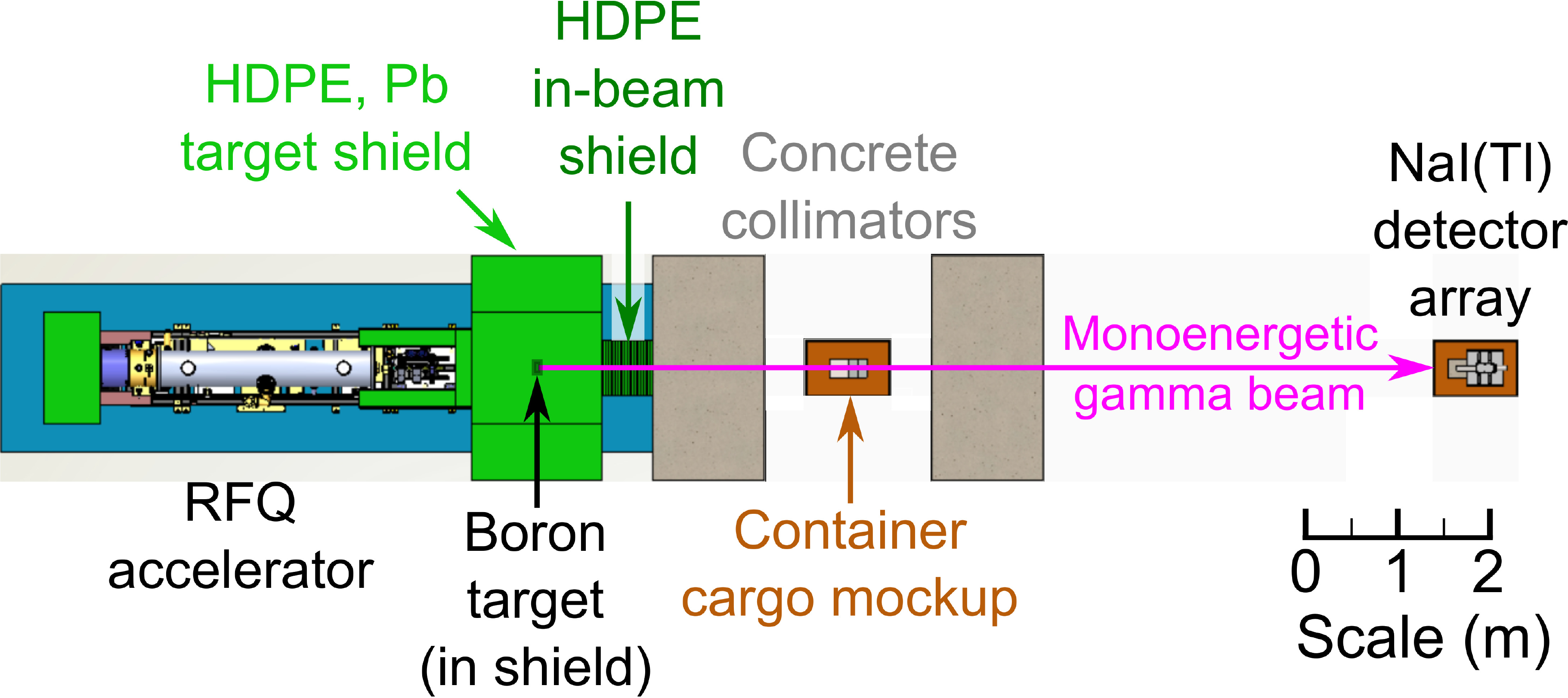}
\caption{Schematic of the mock cargo scanning experiment, viewed from above \cite{buck}.  The arrow
associated with the mock cargo label indicates the direction of motion of the cargo across the
scan.} \label{fig:schem}
\end{figure}

% FIX: Ensure MMGR is defined before this
To test the capability of the MMGR technique,
a mock cargo scanning setup was constructed at the MIT-Bates Research
and Engineering Center, a schematic of which is shown in Figure \ref{fig:schem}.
This setup expanded upon previous test experiments\cite{buck,jill}
to permit the 2D imaging of mock cargo materials.  This included the installation
of a motion system to move mock cargo materials through the beam, an array
of 32 detectors to provide position resolution perpendicular to the direction of the motion,
and the addition of a number of beam and data diagnostics to monitor the system over the
course of a scan.  This section describes the key elements of the experiment and mock
cargo scenario for which data was collected.

\subsection{Gamma Ray Beam}

The 4.4 and 15.1 MeV photons used to radiograph materials were generated by
impinging a 3 MeV deuteron ($d^+$) beam on a thick natural boron target, containing 80.1\% $^{11}$B.  Given the relative cross sections
of the $^{11}$B(d,n$\gamma$)$^{12}$C for the 4.4 and 15.1 MeV gammas at this energy, the beam on target
produced the two gammas in approximately a 4:1 ratio\cite{COOPER201345,CLASS1965433}. The deuterons
were accelerated using an Accsys Technologies DL-3 Radio Frequency Quadruple (RFQ) accelerator,
and the target was mounted to the output port of the RFQ.  The accelerator operated at a frequency
of 300 Hz, producing deuteron beam pulses of approximately 20~{\textmu}s (0.6\% duty factor) as shown
in Figure \ref{fig:pulse}.  Thus, while the time average deuteron current during experiments was approximately
10~{\textmu}A, the instantaneous current during the beam pulses reached $\sim$1.7~mA.

\begin{figure}[thb]
    \centering
    \includegraphics[width=\columnwidth]{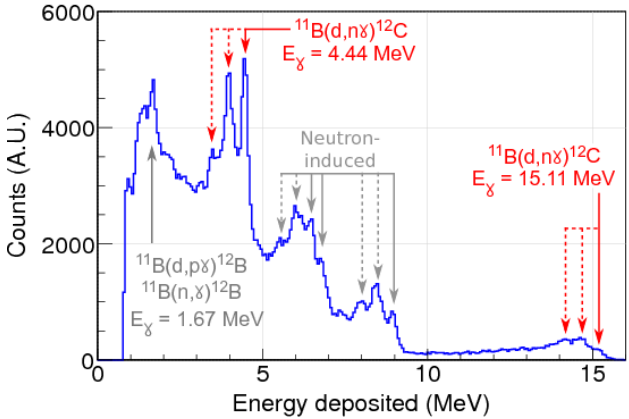}
    \caption{Sample spectrum measured in the NaI (Tl) detectors with an iron sample in the beam
             and $\sim$1~{\textmu}A deuteron current, so as to show the features with high resolution.  See
             O'Day, \textit{et al.}\cite{buck} for a discussion of the labeled elements of the spectrum.}
    \label{fig:slab}
\end{figure}

The reactions at the target generated gamma rays approximately isotropically \cite{CLASS1965433}.  Thus, high-density concrete
collimators were used to create a fan beam extending vertically with an illumination width of 2.38 cm in the horizontal direction
at the location of the mock cargo.  Additionally, 53 cm of borated high density polyethylene (HDPE) was placed directly downstream
of the target (encompassing the entire fan beam) to block neutrons and low energy photons from secondary reactions in the target.
Figure~\ref{fig:slab} shows the spectrum of the beam measured at low event rate to show the key features, including the 4.4 MeV and 15.1 MeV gammas.
An additional contributions is visible at 1.7 MeV (from $^{11}$B(d,p$\gamma$)$^{12}$B
and peaks between 6 and 9 MeV result from thermal neutron capture in the detectors and surrounding materials.
See O'Day, \textit{et al.}\cite{buck} for an extended discussion of the beam components.

\subsection{Detectors}

The transmitted spectra were measured using a vertical array of 32 Saint-Gobain 2X4H16/2SS NaI(Tl) scintillator
detector packages \cite{stg}.  The detectors consisted of $2\,'' \times 4\,'' \times 16\,''$ thallium-doped
sodium iodide crystals instrumented with $2\,''$ photomultiplier tubes.  The large size and appreciable energy
resolution of these detectors allowed the selection of directly transmitted monoenergetic photons, providing critical
information for precision material identification.
The high voltage and gain controls
were manually adjusted for each detector to approximately match their responses, although further energy calibration (gain) corrections
were applied in analysis (see Section \ref{sec:gain}).  The array was constructed so that the
long axis of the detectors was parallel to the beam axis and the short axis was along the vertical direction to
maximize the vertical spatial resolution of the detector array.  The detector array was placed such that the
upstream faces of the detectors were 9.35~m from the boron target (or approximately 5.81~m from the mock cargo).
This resulted in approximately 3~cm vertical resolution for the cargo imaging.  The horizontal extent of the
detectors perpendicular to the beam was wider than the collimation, and thus did not significantly affect the
imaging resolution.

\subsection{Data Acquisition}

The detector pulses were processed using CAEN V1725 digitizer modules operating in digital pulse processing
pulse shape discrimination (DPP-PSD) mode \cite{caen}.  The system was configured such that
the trigger threshold for each detector approximately corresponded to a 1 MeV energy deposition.
The pulse integration window for each trigger was 1 {\textmu}s. Note that unlike standard radiography systems,
which operate in charge integrating mode, the system described here recorded individual
waveforms with timing and pulse shape information available for each detection. This allowed
the use of several analysis techniques described in Section \ref{sec:analysis} to increase
the resolution of the system in effective $Z$ and areal density.  The digitizer output was
processed using an extension to the ADAQ analysis framework to produce data files for analysis \cite{adaq}.

\subsection{Mock Cargo Test Configuration}
\label{sec:mock}

To utilize this system as a cargo scanning prototype, a motion system installed between the concrete collimators
(as shown in Figure~\ref{fig:schem}) moved materials samples placed on a cart (shown in Figure~\ref{fig:images})
across the fan beam over the course of an experimental run.   Data were collected as a function of time, which,
when paired with the known motion of the materials and the vertical resolution of the detector
array, allowed the 2D imaging of the mock cargo materials.  The materials tested were chosen to so as to span a large
range of effective $Z$ ($\sim$5--82) and to include a stand-in for SNM (natural uranium rods with
aluminum cladding---see Appendix~\ref{app:eff}).  The areal densities of the materials were chosen so as to approximate
typical total areal densities present in commercial cargo containers.  Table~\ref{tab:mats} summarizes the parameters of the materials
samples.

Section~\ref{sec:res} presents results from two distinct experimental runs using the same materials samples: one in which
the cargo was moved across the beam at 0.0077 cm/s and one in which the cargo was moved at 0.308 (4x the speed of the first test).
These are referred to as the 7400~s and 2000~s scans respectively.  Additionally, the data could be sampled to considerably
finer time resolution (4 ns), permitting the oversampling of the data relative to the collimator width to improve the horizontal
position resolution of the reconstruction.  Analyses were conducted using 1 cm and 1 mm pixel widths, as discussed in Sections~\ref{sec:analysis}
and \ref{sec:res}.  Note that while these scan times are considerably longer than would be feasible for a deployed cargo scanner, the
relevant quantity is the integrated deuteron beam current delivered on target per unit scan distance, since scan times may be reduced by increasing the beam
current.  In the 7400~s run, the beam charge delivered was 1.3~mC/cm of scan length (at 10 {\textmu}A of average beam current).  This would correspond
to scan times of $\sim$100~s for full sized containers with 1~mA average beam current.  Such currents would likely be achievable
using a purpose-designed accelerator (operating with continuous wave current).

\begin{table*}[thb!]
\begin{center}
\begin{tabular}{l|r|r|c|r|r}
Material & Effective $Z$ & Density  & Width $\times$ Height  & Depth  & Areal Density ($\rho_A$) \\
         &   &  (g/cm$^3$) &  (cm$\times$cm) & (cm)  &  (g/cm$^2$) \\
\hline\hline
Borated HDPE  & $\sim$5.2 & 1.02    & $20.3\times23.1$ & 45.30 & 46.2 \\
Aluminum (Al) & 13        & 2.70    & $20.3\times24.5$ & 20.25 & 54.7 \\
Copper (Cu)   & 29        & 8.96    & $10.1\times10.1$ &  5.45 & 48.8 \\
Tin (Sn)      & 50        & 7.31    & $10.1\times10.1$ &  6.74 & 49.3 \\
Tungsten (W)  & 74        & 19.30   & $10.1\times10.1$ &  2.56 & 49.4 \\
Lead (Pb)     & 82        & 11.35   & $20.3\times20.3$ &  5.08 & 57.7 \\ 
Uranium rods & $\sim$65  & 12.72   & $13.8\times21.2$ &  5.50 & $\sim$55\\ % FIX (appendix note)

\end{tabular}
\end{center}
\caption{Parameters of the materials samples used for the imaging test, the arrangement of which is shown in Figure~\ref{fig:images}.  The effective $Z$ value listed for the borated HDPE
is computed as the average elemental composition of the material weighted by the contribution to the electron density by
each element, since Compton scattering dominates the photon interactions at the energies of interest for the light nuclei
comprising the material \cite{xcom}.  Values listed for the uranium rods
are averaged over the arrangement of the 10 rods.  See Appendix~\ref{app:eff}
for explanation of the effective $Z$ and areal density for the uranium rods.}
\label{tab:mats}
\end{table*}

\section{Analysis}
\label{sec:analysis}

To reconstruct the effective $Z$ and areal density of the mock cargo, the
transmitted gamma ray spectra
were compared to the expected spectra based on a detailed simulation model of
the experiment.  The data spectra were collected over fixed increments of the scan
length for each detector channel to create ``pixels'' for the material reconstruction.
Similarly to standard radiography techniques, the analysis
consisted of comparing the transmitted spectra with materials in the beam to
that of the ``open'' beam, i.e., when no materials were present in the bean other than the fixed components of the setup described
in Section~\ref{sec:exp}.  This comparison provides information of the
total attenuation of the beam due to the material as well % CHANGE
as the energy dependence of this attenuation, which provides sufficient information
to reconstruct the total areal density and effective atomic number of the materials.

The simulation model was used to generate a library of materials
over the complete space of $Z=4\text{--}92$ and areal density $\rho_A=20\text{--}250$~g/cm$^2$. % CHANGE
The use of the simulation library allowed for the reconstruction of the cargo materials
without empirical calibration based on additional datasets and provided a means of
directly accounting for detector response and efficiency, collimation of the beam, 
multiple/down-scattering of transmitted photons, and
other elements of the physical setup.  This section describes the procedures applied to prepare
the data spectra for comparison with the simulated transmission library, the simulation model, and the 
analysis used to extract the effective $Z$ and areal density of the mock cargo.

\subsection{Spectrum Corrections}
\label{sec:corr}

Several corrections were applied to the raw spectra, both to ensure consistency
across an imaging scan and to select the relevant data for comparison with the
simulation model.
Unlike traditional gamma/x-ray cargo radiography systems, which utilize
integration mode detectors to cope with the high photon flux of bremsstrahlung
beams \cite{r60}, the lower absolute photon flux of the nuclear reaction based
photon beam here permits use of detectors in counting
mode. This makes it possible to record the complete energy dependence of the transmitted spectra.
This spectral information allows individual recorded events to be associated with the
initial photon energy and thus more accurate determination of
the attenuation of the beam due to the cargo at the specific beam energies.
To produce spectra representative of the transmission of the monoenergetic photons,
several corrections must be applied to the raw spectra.  The corrections are described
as follows in the order they were applied to the raw data.

\subsubsection{Gain Drift Correction}
\label{sec:gain}

As the experiment operated in a non-climate controlled warehouse, the NaI (Tl)
detectors were subject to gain drift on the order of several percent
over the course of each scan.  Since the analysis depends
on the measurement of counts recorded in specific energy regions of the data spectra,
a fixed calibration of ADC counts to deposited energy for a detector would cause
systematic error for each energy bin.  To prevent this, the raw spectrum of each
detector at each position step in the scan was used to determine the ADC-to-energy
calibration at that specific step using the monoenergetic peaks present in the spectra
to produce energy spectra that could be compared on equal footing.

\subsubsection{Beam Timing Cut}

The pulsed nature of the deuteron beam provided a means of suppressing many of the background
contributions to the raw detected spectra.  Since the lifetimes of the excited states of $^{12}$C
that gave rise to the 4.4 and 15.1 MeV photons in the target are $\mathcal{O}\left(10^{-13}\:\text{s}\right)$,
events from the gamma rays of interest were recorded promptly in coincidence with beam pulses.  Gating on the beam pulse
timing allowed for suppression of background events due to longer-lived excited states and thermal neutrons.
In particular, bremsstrahlung
photons arising from the beta decay of $^{12}$B (produced by neutron capture on $^{11}$B in the target) contributed
significantly to the raw signal up to 6.9 MeV.  Given that the beta decay of $^{12}$B has a
lifetime of $\sim$20 ms, however, $\sim$99.9\% of its contribution to the raw spectra may be eliminated
by selecting only events in the beam pulse time windows.

\begin{figure}[thb]
    \centering
    \includegraphics[width=\columnwidth]{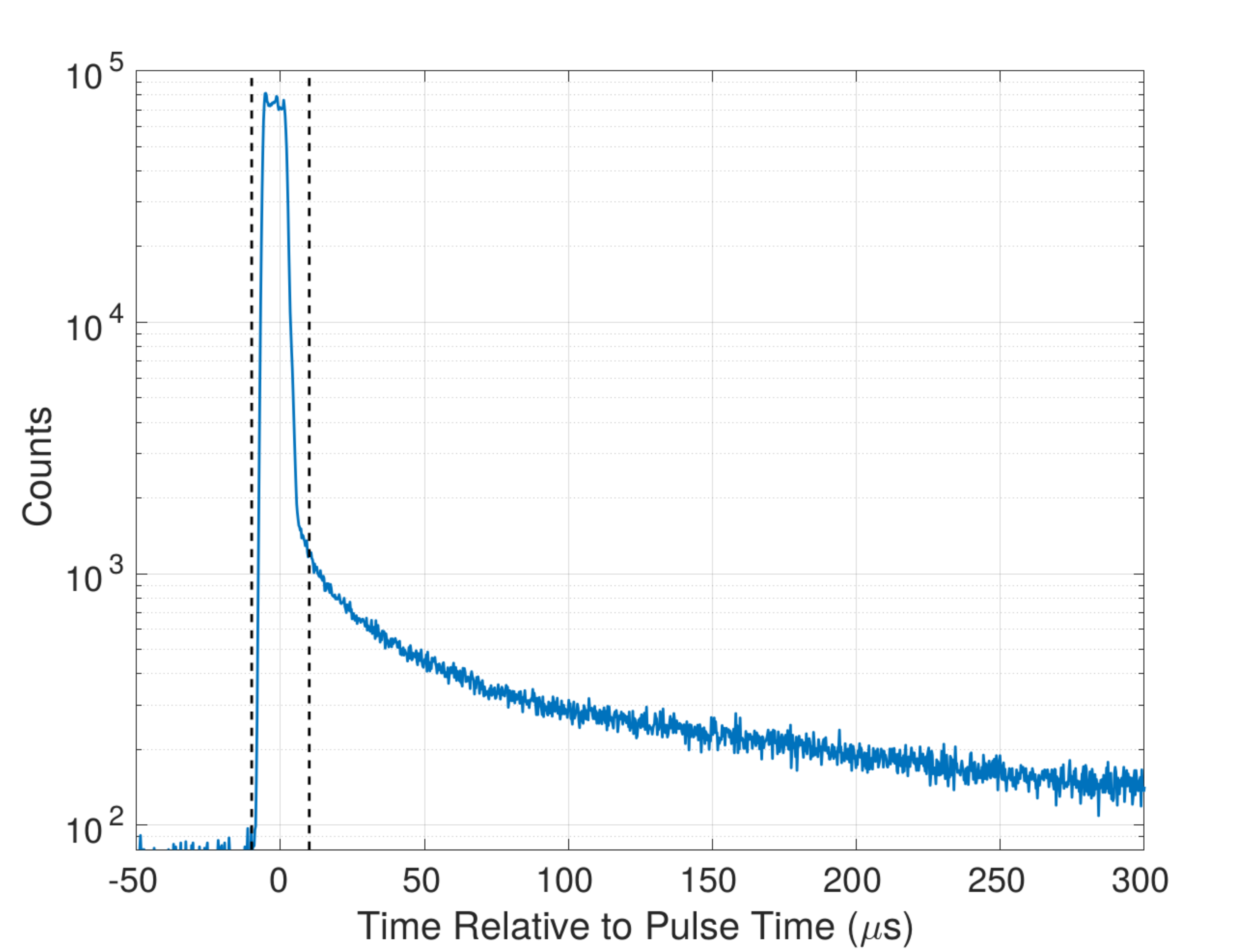}
    \caption{Reconstructed beam pulse shape, time behavior of the after pulse events, and cut window
             applied to select only prompt events associated with beam pulses (dashed lines).}
    \label{fig:pulse}
\end{figure}

\begin{figure}[thb]
    \centering
    \includegraphics[width=\columnwidth]{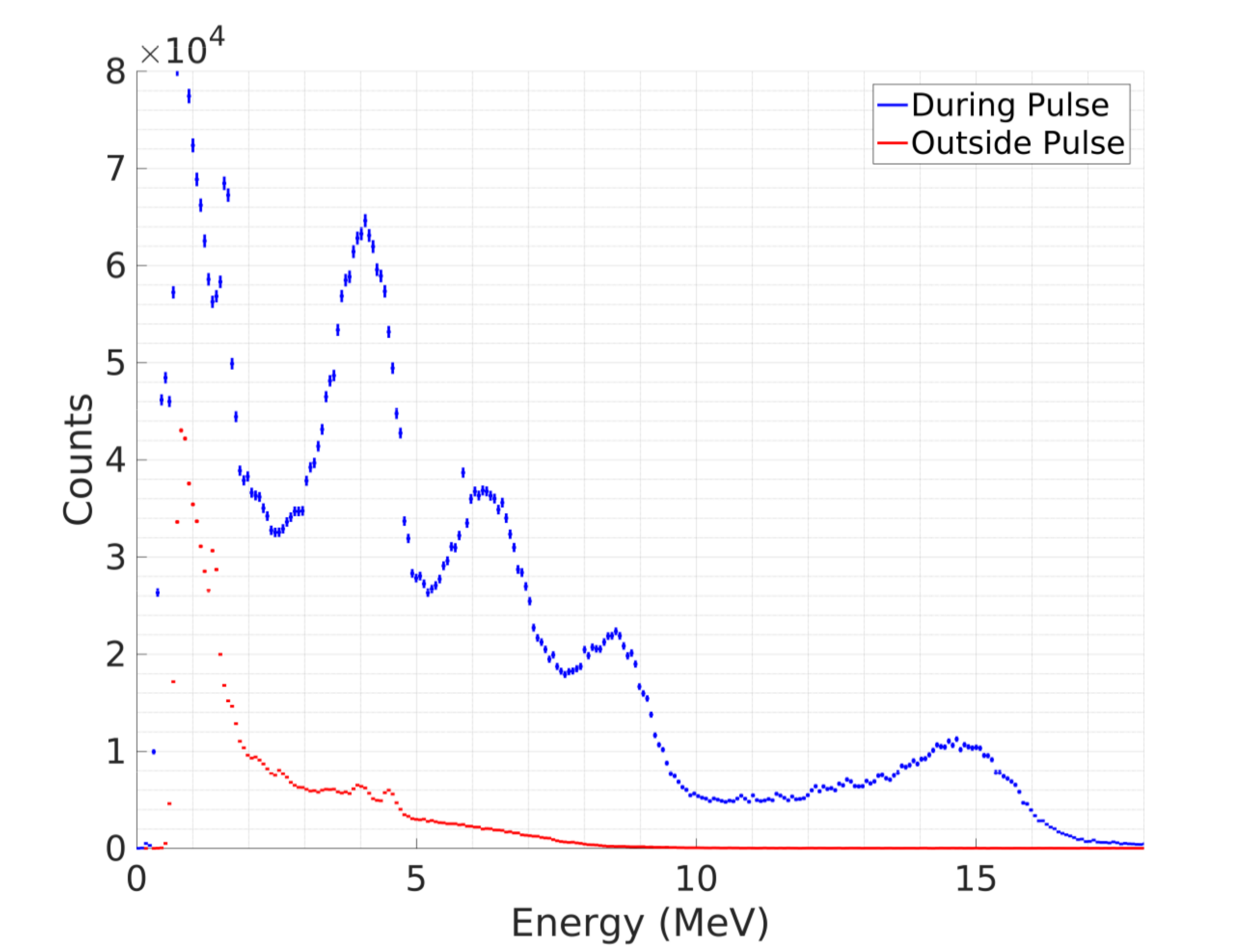}
    \caption{Spectra of the open beam during and outside the beam pulse window (lower histogram, color online), with
             pile-up corrections applied (see Section~\ref{sec:pup}).  Events outside the pulses contributed approximately
             13\% of all raw counts, after correction for pile-up (Section \ref{sec:pup}).}
    \label{fig:timec}
\end{figure}

Since no timing information was recorded for the beam pulses during data taking and the exact frequency
of the accelerator deviated slightly from 300 Hz, the pulse frequency
was reconstructed by computing the mean time of concentrations of events in the detectors over many pulses.
A symmetric 20 {\textmu}s window around the reconstructed pulse center was selected for the timing cut, as shown in
Figure~\ref{fig:pulse}.  Figure~\ref{fig:timec} shows the spectrum of the open beam inside and outside
the timing cut, showing that the inclusion of events outside the time cut would contribute $\sim$10\%
error to the estimated counts in the 4.4 MeV region.  Notably, the 4.4 MeV signal remains visible in the off-pulse
spectrum.  This is due to the fact that the beta decay of $^{12}$B frequently creates the $^{12}$C 4.4 MeV excited
state \cite{b12b,a12}.  Additionally visible are small peaks from the capture of thermal neutrons on hydrogen (2.2 MeV)
and a longer lived excited state of $^{12}$B (1.7 MeV)~\cite{a11}.

\subsubsection{Beam Current Correction}

Any unaccounted for variation in the deuteron current
(and thus the beam flux) between the open beam and the subsequent measurements would caused an error in the
transmission measurement directly proportional to the current variation.  As noted in Section~\ref{sec:exp}, a charge integrator was used to monitor the beam current incident
on the boron target over the course of each imaging scan.  The beam current varied by up to $\sim$10\% during the imaging
tests, primarily due to instabilities in the deuteron source and accelerator.
The data from this channel were used to renormalize the data spectra at each position step.
Note that only the relative beam current at each scan step is required, rather than an absolute calibration, since the analysis
utilizes only the relative transmission between cargo-in-beam and the open beam.  While an approximate calibration of the current
was known, any uncertainty in its value does not significantly affect the reconstruction of the materials.

% \begin{figure}[thb]
%     \centering
%     \includegraphics[width=\columnwidth]{figures/beamc.pdf}
%     \caption{Measurement of the deuteron beam current incident on the boron target over the course of the two imaging scans.}
%     \label{fig:beamc}
% \end{figure}

\subsubsection{Pile-Up/Dead-Time Correction}
\label{sec:pup}

The large size of the NaI (Tl) detectors in combination with the high instantaneous current
of the pulsed RFQ beam resulted in a significant number of ``pile-up'' events in the raw spectra
(i.e., single spectrum counts representing the energy deposition of two or more individual photons
in the same pulse integration period).  These pile-up events significantly distort the open beam
spectrum. In particular, such events add an excess of events at higher energies from the summation of two
lower energy depositions.  While standard radiography systems operating with integration mode
detectors are not subject to this issue, the analysis described here --- which utilizes spectral information --- must
carefully account for this effect.

A pulse shape discrimination (PSD) algorithm was used
to identify such pile-up events in data.  The ``tail-over-total'' PSD method, frequently
used to separate gamma ray and neutron events in organic scintillators due
to their differing scintillation decay time scales \cite{psd}, may also be used to
identify pile-up events.  In this method, the charge integrated by the ADC for 
a PMT waveform is separated into ``head'' and ``tail'' portions at a
a fixed amount following the trigger.  An energy deposition of any value resulting from
a single event should exhibit roughly the same ``tail-over-total'' ratio, corresponding to
the decay time of the scintillator.  Integration windows with pile-up events
will show an excess in the tail portion of pulse integration due to the contribution of the second pulse.

\begin{figure}[thb]
    \centering
    \includegraphics[width=\columnwidth]{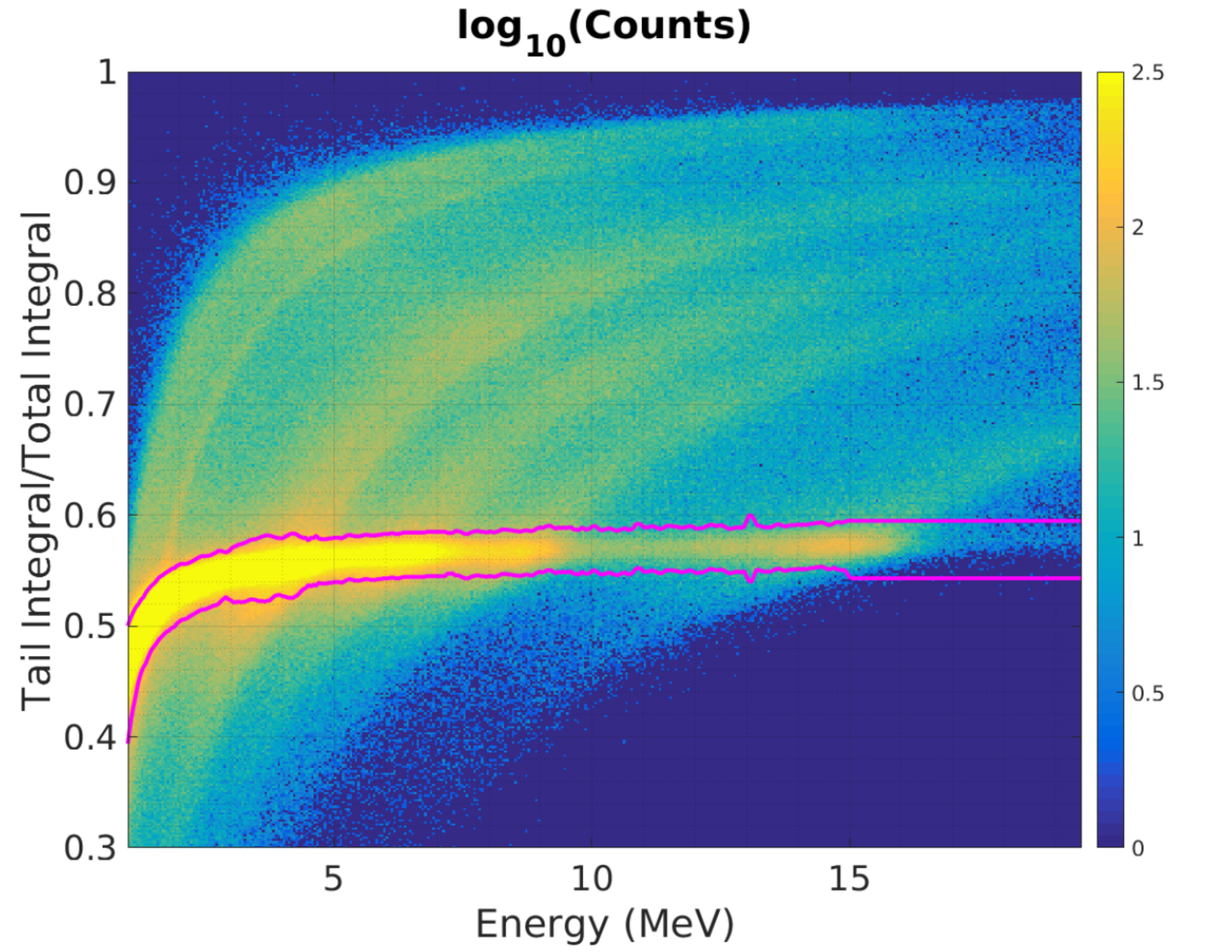}
    \caption{Two-dimensional ``tail-over-total'' PSD histogram ($\log_{10}$ counts) for the raw
             spectrum of the open beam.  The bright band at a ratio of $\sim$0.55 represents
             windows with a single detected photon. The curved bands away from the main band show pile-up events,
             in which additional energy is added to the common monoenergetic depositions (above the main band) or in which
             a monoenergetic event occurs close enough to the end of a trigger window for its tail to cause another
             trigger (below the main band).  The region
             between the magenta lines indicates the pile-up cut region.}
    \label{fig:psd}
\end{figure}

For the imaging scans, the pulse integration period was fixed at 1 {\textmu}s, and the tail region
was defined to be approximately the last 50\% of the pulse following the trigger.  The 2D histogram
of the tail fraction and the total energy deposition for a single detector over the course of
one of the data runs is shown in Figure~\ref{fig:psd}.  For each energy bin in each detector, a Gaussian profile
was fit to the central tail/total ratio peak to produce a cut at $3\sigma$ for each
energy bin in each detector to reject events as pile-up.  The resulting cut
region on the PSD parameter for an example detector is also shown in Figure~\ref{fig:psd}.

% FIX: Maybe say something about the ``bend'' in the figure below, maybe leave it to be asked if needed

Rejecting events identified as pile-up, however, introduces an effective deadtime to the measurement
(since no counts are accepted in an integration window with pile-up).  Since the presence of material
in the beam significantly affects the pile-up rate, the pile-up rejected spectra on their own are
not representative of the actual transmitted flux compared to the open beam.
For the open beam, approximately 70\% of recorded energy
depositions were rejected due to pile-up (after correction for the Gaussian cut boundaries).
Since each of these pile-up events represents at least two
individual photons, use of the uncorrected pile-up rejected spectrum would result in underestimating the true
flux by $>$500\%, and would introduce similarly large errors in the transmission ratio.  Due to the long
integration window, pile-up deadtime dominated the total effective deadtime. Additional data acquisition
and processing time added $\ll$1\% to the effective deadtime, and was not included as a significant correction.

To account for this, a pile-up correction was
devised.  Given that individual true events are independent, the true energy spectrum
of counts that are recorded in the pile-up portion of the data matches
the energy spectrum of the counts that occurred without pile-up, up to secondary
effects such as small additional energy pile-up and associated trigger bias. Thus the
desired correction may be approximated as a scaling factor that is applied to the pile-up
rejected (``clean'') spectrum.  Since the true rate of individual events $r$ is unknown, and
the resolving time of the detector $\tau$ may also be unknown, it is most useful
to express the standard formulation of pile-up \cite{knoll} as a function of the fraction $f$ of total counts captured
in the pile-up rejected spectrum.  For the long, fixed ADC integration window used in this experiment, the
deadtime was was of a non-paralyzable nature.  As derived in Appendix~\ref{app:pcor},   % CHANGE (REWORDED)
the true spectrum $N$ may thus be reconstructed from the pile-up rejected spectrum $N_C$ as:
\begin{equation}
 \label{eq:pcor}
 N = N_C\left( \frac{1-\ln f}{f} \right).
\end{equation}
This correction was applied to each time/position step of the image scans to account
for variation in the pile-up rate over the course of the experiments.

\subsection{Simulation Model}

A complete simulation model of the experiment was constructed, including all relevant aspects of the experiment,
to compute simulated transmission spectra for a wide variety of materials
that could be directly compared to the data.  By using such a model, the need for empirical
calibration of the system with sample data from many materials was avoided.  The simulation model was constructed using the Geant4 toolkit \cite{geant},
and included all important physical materials present in the experiment (the neutron shield in the
beamline, the collimators, the detectors, etc.) at positions surveyed during data taking.  Photons originating
at the boron target were propagated through the geometry (including any simulated cargo material) to the
model detectors,  which included simulated responses --- resolution, efficiency, etc. --- modeled according
to dedicated empirical tests.

\begin{figure}[thb]
    \centering
    \includegraphics[width=\columnwidth]{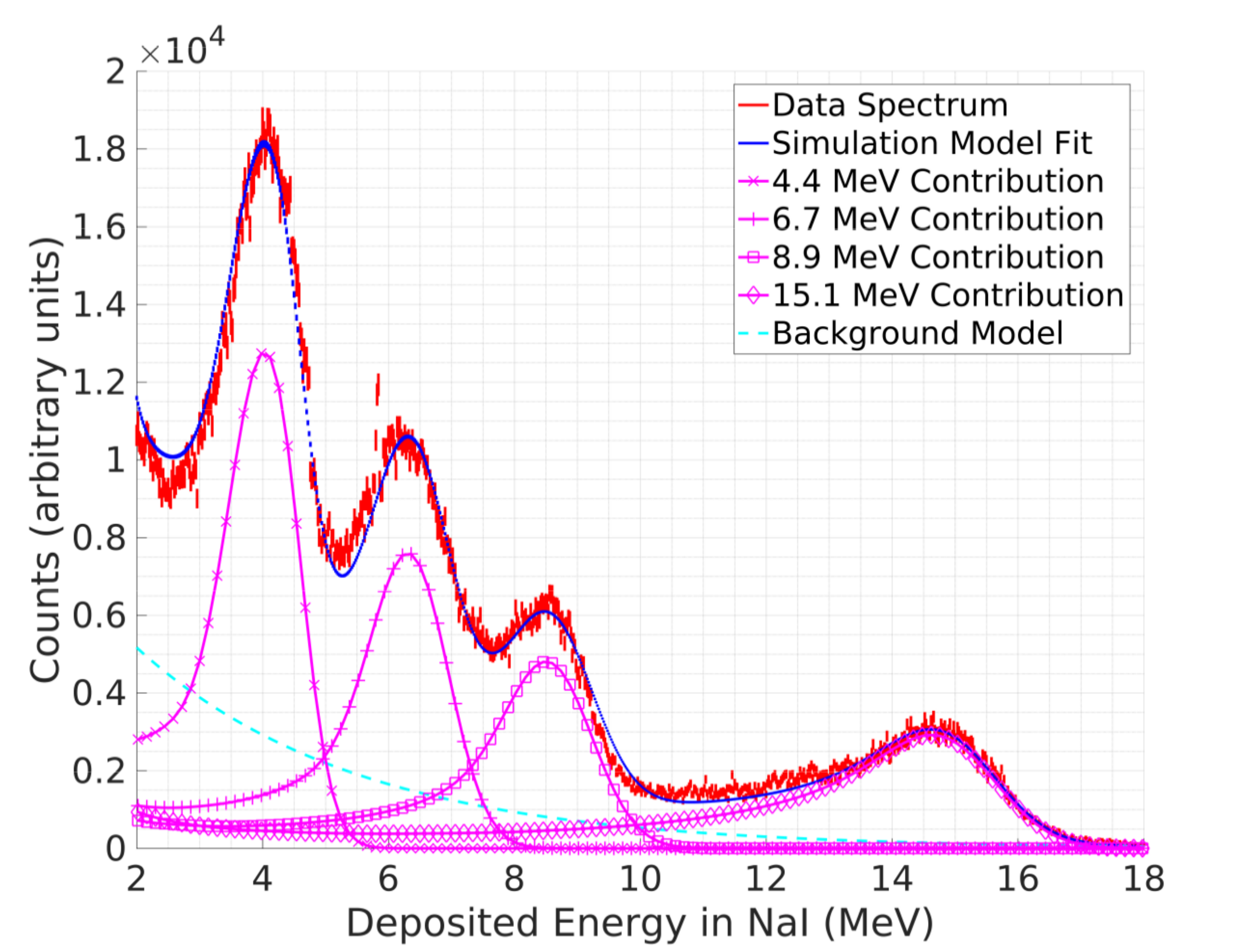}
    \caption{The simulation spectrum fit to an open beam data sample taken using a single NaI (Tl) detector
            (color online).  The individual simulated contribution magnitudes, detector model parameters, and background
            model were fit to best match the data for each sample.}
    \label{fig:smatch}
\end{figure}

\begin{figure}[thb]
    \centering
    \includegraphics[width=\columnwidth]{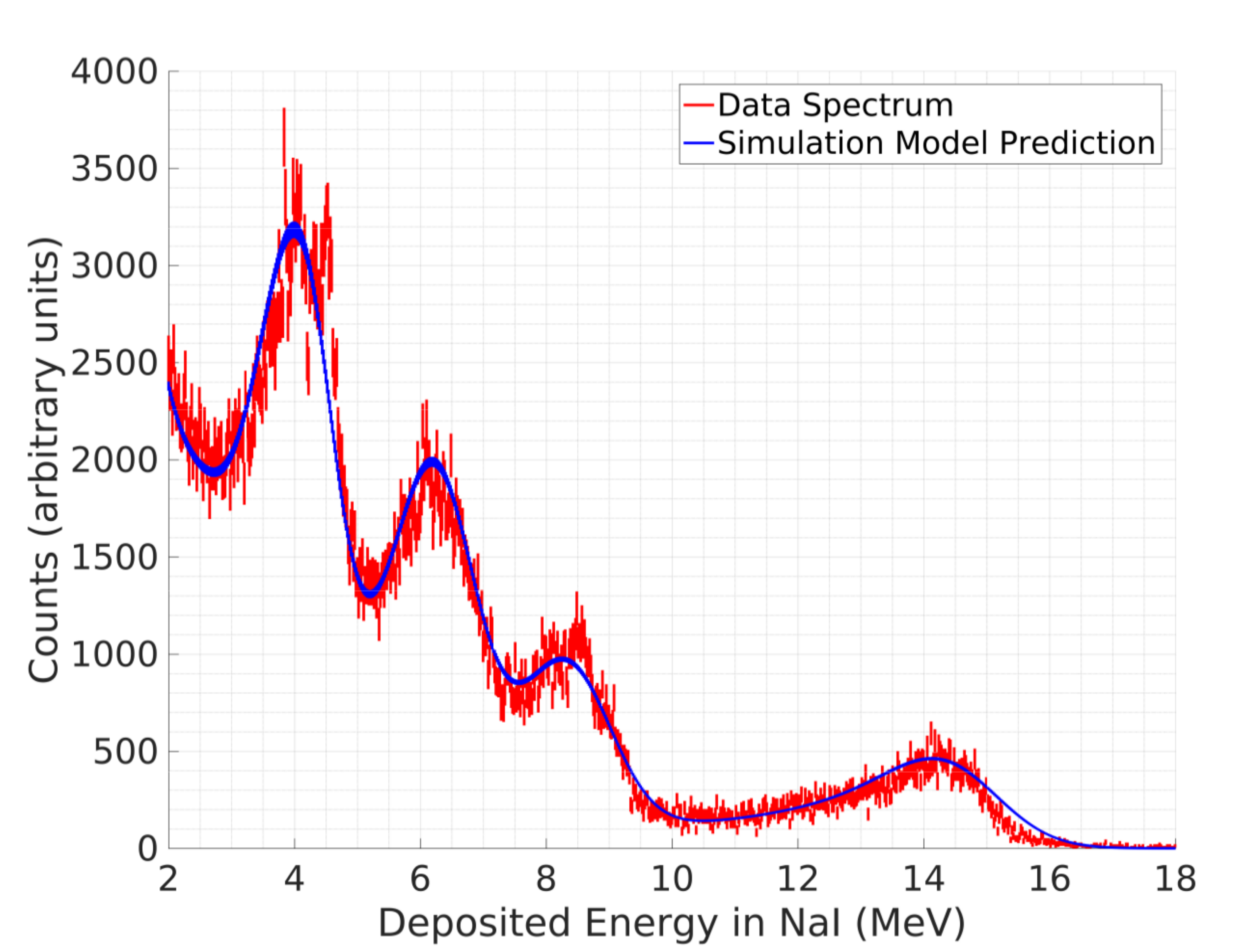}
    \caption{Comparison of the data spectrum from a pixel in the copper region of the 2000~s scan to the simulation prediction based on the fit to the
             open beam spectrum of Figure~\ref{fig:smatch}, given knowledge of the material in the beam at the time.  }
    \label{fig:cucomp}
\end{figure}

The simulated beam was generated by simulating the five major monoenergetic beam components
shown in Figure~\ref{fig:slab} (1.7, 4.4, 6.7, 8.9, and 15.1 MeV) for fixed geometries, and determining their
relative contribution to the beam using an empirical fit to data taken in dedicated experiments.  The resulting simulated spectra for each monoenergetic
contribution were convolved with the simulated detector response, and then the relative strength of each
contribution was fit so as to best match the corresponding data.  The results of the fit for the open beam is shown
in Figure~\ref{fig:smatch}.  Note that while the analysis depends only on relative transmission, this beam model
accounts for the contributions in 4.4 MeV region due to downscatter and incomplete energy deposition in the detectors from higher energy photons
to increase the accuracy of the analysis.  Figure~\ref{fig:cucomp} shows the simulation prediction for the transmission spectrum for a copper sample of $\rho_A \approx 49$ g/cm$^2$.
This prediction is based on propagation of the reconstructed open beam (Figure~\ref{fig:smatch}) through a simulated material sample
using Geant4.  These predicted spectra were compared to the data spectra to reconstruct materials, as described in Section~\ref{sec:recon}.

Approximately 5000 simulated transmission experiments were generated to create a library
of expected detected spectra over the complete two-dimensional space of $Z= 4\text{--}92$ and 
$\rho_A = 20\text{--}250$ g/cm$^2$, in addition to the open beam configuration.  As described in Section~\ref{sec:recon}, % CHANGE
this simulation library was compared to the transmission data to estimate the areal density and effective $Z$ of each
pixel in a scan.

\subsection{Radiographic Reconstruction}
\label{sec:recon}

With data and simulation spectra prepared, the transmission ratios for each data and simulation spectrum
were computed in the regions of the 4.4 MeV and 15.1 MeV peaks.  For each spectrum (with the
simulation and data treated in the same manner), the
counts between 2.8 MeV and 5.0 MeV (so as to encompass the full deposition and escape peaks of the
4.4 MeV photons) were integrated to compute $N_4$, while all counts above 10.1 MeV were integrated to
produce $N_{15}$ (since essentially all counts above this energy were due to incomplete energy depositions
of 15.1 MeV photons).  The simulation was run with no mock cargo to produce an open beam spectrum for the
transmission calculation, while the data runs used the open beam spectra collected during the first and
last portions of the run (normalized to the integrated current of one pixel).  The transmission ratios
in each energy bin $E$ were defined as
\begin{equation}
 R_E = \frac{N_E}{N_{E,\text{open}}},
 \label{eq:trat}
\end{equation}
where the $N_{E,\text{open}}$ are the integrated counts for the appropriate open
beam spectra, after applying the corrections discussed in Section \ref{sec:corr}.

\begin{figure*}[thb]
    \centering
    \includegraphics[width=\columnwidth]{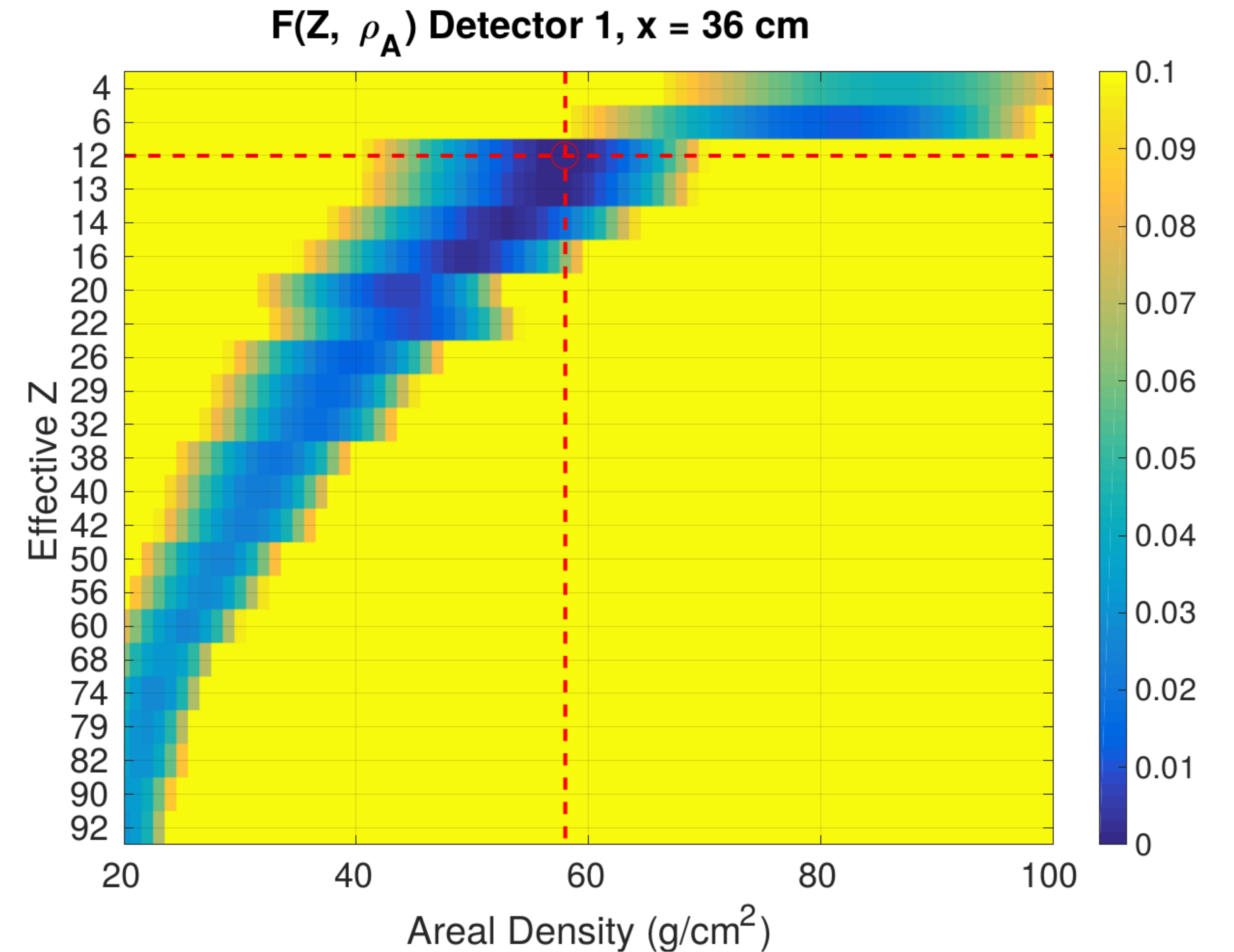}
    \includegraphics[width=\columnwidth]{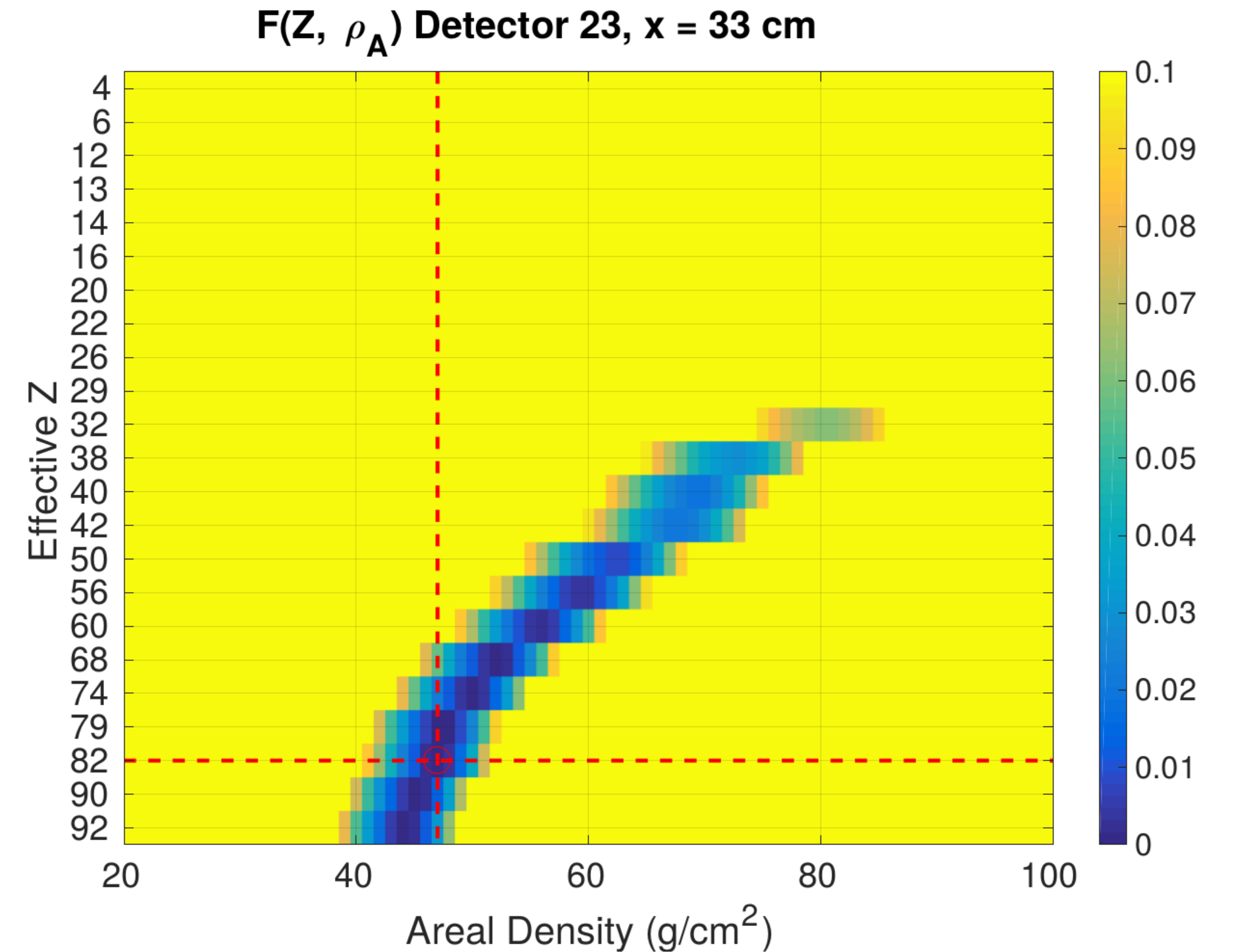}
    \caption{Two examples of the data/simulation comparison metric $F$ (Equation~\ref{eq:metric}) as a function of $Z$ and $\rho_A$ over the simulation
    library.  Examples from the aluminum (left) and
    tungsten (right) regions of the 7400~s test are shown.  The crossing dashed lines indicate the minimum of $F$ in each plot, showing the reconstructed
    $Z$ and $\rho_A$ values.}
    \label{fig:fmin}
\end{figure*}

With the transmission ratios for the data and simulation spectra in the regions of interest determined,
a figure of merit $F$ was constructed to determine the simulated combination of $Z$ and $\rho_A$
that best matched the data spectrum.  The quantity was constructed using ratios of the 
transmission ratios (Equation \ref{eq:trat}) in the 4.4 and 15.1 MeV regions to construct a quantity robust
against a number of systematic uncertainties.  For each pixel, the data spectrum was compared to
each element of the simulated material library with effective $Z$ and $\rho_A$ according to
\begin{multline}
 \label{eq:metric}
 F\left(Z,\rho_A\right) = \left( \frac{R_{15,\text{data}}}{R_{15,\text{sim}}\left(Z,\rho_A\right)} - 1 \right)^2 \\
    + \left( \frac{ \sfrac{R_{4,\text{data}}}{R_{15,\text{data}}} } {\sfrac{R_{4,\text{sim}}\left(Z,\rho_A\right)}{R_{15,\text{sim}}\left(Z,\rho_A\right)}}  - 1 \right)^2.
\end{multline}
This metric was motivated by the fact that the signal in the 15.1~MeV region was very
clean due to the absence of high energy backgrounds in the data and the fact that the ratio-of-ratios
between the 4.4 and 15.1~MeV regions provides strong material discrimination while canceling certain
systematic uncertainties.
The values of $Z$ and $\rho_A$ corresponding to the minimum of $F$ were assigned as the
reconstructed values for each pixel, noting that a perfect match between the measured
and simulated transmission results in $F=0$.  Figure~\ref{fig:fmin} shows examples
of the reconstruction for a low-$Z$ material (Al) and a high-$Z$ material (W).

\section{Results}
\label{sec:res}

\begin{figure*}[thb]
    \centering
    \includegraphics[width=2.1\columnwidth]{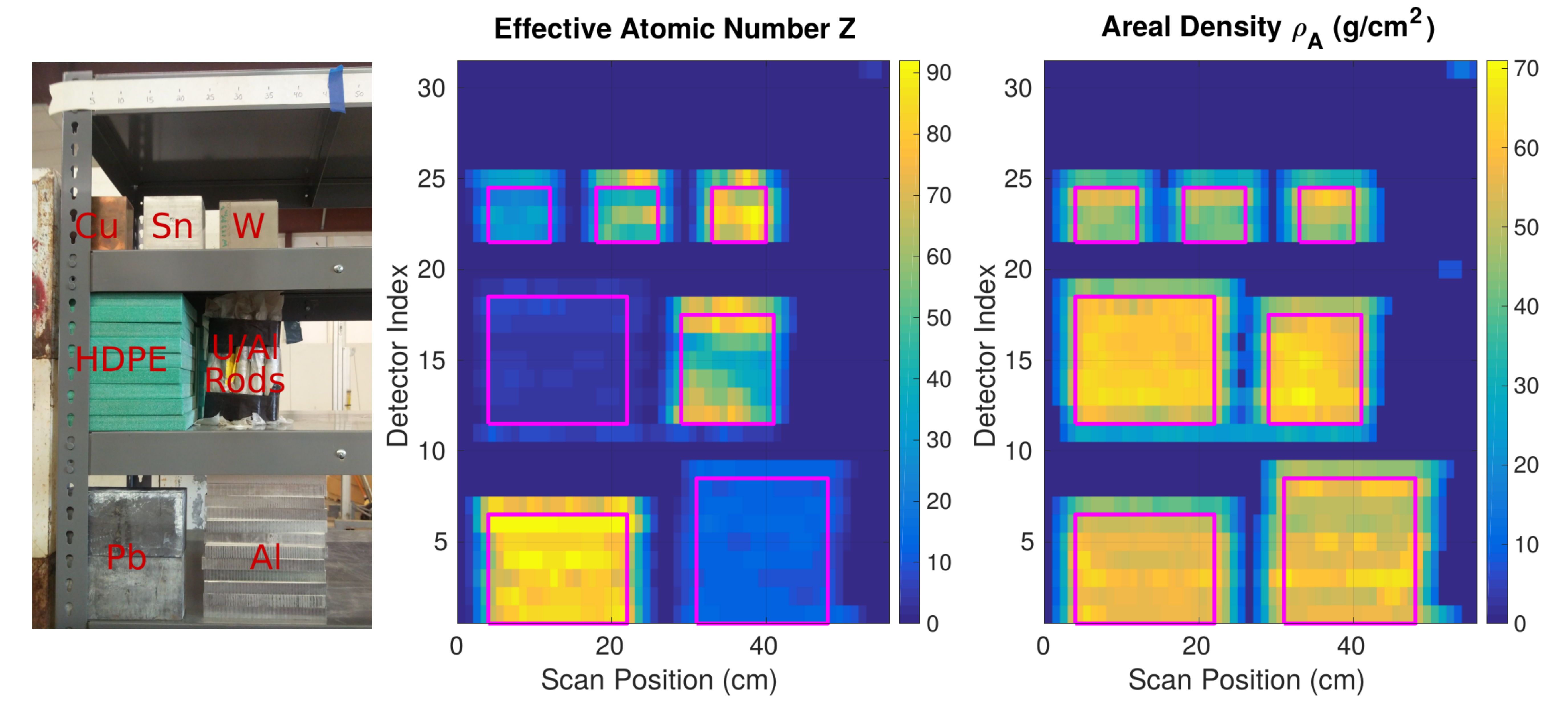}
    \caption{Images of the mock cargo in reconstructed effective $Z$ and areal density for the 7400~s test ($\sim$1.3 mC integrated deuteron beam current on target
per cm of the scan).  The magenta boxes show the regions used to define each material sample for the computation of the values in Table~\ref{tab:long}. }
    \label{fig:images}
\end{figure*}

To produce images of the mock cargo in effective $Z$ and areal density, the data spectra
were grouped into 1 cm pixels along the scan length.  For each pixel, the radiographic
analysis described in Section~\ref{sec:analysis} was applied to produce the images in
effective $Z$ and areal density $\rho_A$. Figure~\ref{fig:images} shows the results for the 7400~s
scan.  In addition to the pixel-by-pixel estimate, regions were defined according the boxes
shown in the reconstruction images to quantitatively evaluate the performance of the analysis 
for each material.  For each region, the average reconstructed $Z$ and $\rho_A$ were computed
for comparison to the known true values.  These results are summarized in Tables~\ref{tab:long}
and \ref{tab:short} for the 7400~s and 2000~s scans, respectively. The standard deviation of each estimated pixel value
was used as a measure of the uncertainty on the values reconstructed for each individual pixel, which were then
used to compute the uncertainties on the overall mean reconstructed values.

The reconstructed values are very close to the true values in both of the imaging tests, showing the robustness of
the system and analysis to environmental drift effects (e.g., temperature changes) and the statistics of the data (the 
7400~s run included approximately 4 times as many counts in the transmission 
spectra per pixel as the 2000~s run).  It should also be noted that while the 
absolute values are close to the true values, there are residual statistically significant differences.  This indicates that further improvements 
to the reconstruction algorithms or control of systematic uncertainties are possible, and thus should be part of future 
work.  Despite this limitation, the specificity of the monoenergetic
beam transmission provides effective atomic number identification with specificity of $\pm$3 in $Z$ and thus permits separation of different high-$Z$ materials, which is 
typically not possible in
existing radiography systems.  For example, the tungsten and lead samples are well separated in reconstructed atomic
number, while the areal density reconstruction is also accurate for each material to within a few g/cm$^2$.  This suggests that pure special
nuclear materials ($Z\geq92$) could be separated from benign high-$Z$ materials such as lead and tungsten, which would
be invaluable for reducing false alarms in a system designed to detect nuclear smuggling.

\begin{table*}[thb!]
\setlength{\tabcolsep}{6pt}
\begin{center}
\begin{tabular}{l |r r r | r r r}
Material & Actual $Z$ & Reconstructed $Z$ & Single Pixel Unc. & Actual $\rho_A$ & Reconstructed $\rho_A$ & Single Pixel Unc. \\
         &               &  &  & (g/cm$^2$) & (g/cm$^2$)     &    (g/cm$^2$)       \\
\hline\hline
Borated HDPE  & $\sim$5.2 & $4.7\pm0.1$ & $(1.4)$ & $46.2$ & $60.5\pm0.5$ & $(5.3)$ \\
Aluminum (Al) & 13        & $12.0\pm0.2$ & $(2.1)$ & $54.7$ & $55.8\pm0.6$ & $(7.2)$ \\
Copper (Cu)   & 29        & $27.5\pm1.1$ & $(5.6)$ & $48.8$ & $48.0\pm1.2$ & $(6.0)$ \\
Tin (Sn)      & 50        & $49.1\pm2.9$ & $(14.0)$ & $49.3$ & $45.0\pm1.3$ & $(6.2)$ \\
Tungsten (W)  & 74        & $75.3\pm3.3$ & $(15.1)$ & $49.4$ & $50.0\pm1.7$ & $(8.0)$ \\
Lead (Pb)     & 82        & $83.2\pm0.9$ & $(9.3)$ & $57.7$ & $56.8\pm0.4$ & $(4.5)$ \\
Uranium rods  & $\sim$65  & $55.5\pm2.3$ & $(19.3)$ & $\sim$55 & $61.3\pm0.8$ & $(7.0)$ \\

\end{tabular}
\end{center}
\caption{Reconstructed effective $Z$ and areal density values for the mock cargo materials for the 7400~s test ($\sim$1.3 mC integrated deuteron beam current on target
per cm of the scan).  Quoted uncertainty after the ``$\pm$'' represents the uncertainty on the average over the material region, while the
single pixel uncertainty is the standard deviation of the single pixel (1~cm$\times$1 detector) estimates over the region.  See Figure~\ref{fig:images}
for the definition of each sample region.}
\label{tab:long}
\end{table*}

\begin{table*}[htb!]
\setlength{\tabcolsep}{6pt}
\begin{center}
\begin{tabular}{l |r r r | r r r}
Material & Actual $Z$ & Reconstructed $Z$  & Single Pixel Unc. & Actual $\rho_A$ & Reconstructed $\rho_A$  & Single Pixel Unc. \\
         &               &  &  & (g/cm$^2$) & (g/cm$^2$)     &    (g/cm$^2$)       \\
\hline\hline
Borated HDPE  & $\sim$5.2 & $5.9\pm0.2$ & $(2.4)$ & $46.2$ & $61.7\pm0.7$ & $(7.9)$ \\
Aluminum (Al) & 13        & $12.4\pm0.4$ & $(4.6)$ & $54.7$ & $53.8\pm0.9$ & $(10.0)$ \\
Copper (Cu)   & 29        & $28.2\pm1.3$ & $(6.6)$ & $48.8$ & $47.8\pm1.2$ & $(5.9)$ \\
Tin (Sn)      & 50        & $49.7\pm3.3$ & $(16.1)$ & $49.3$ & $46.5\pm1.5$ & $(7.1)$ \\
Tungsten (W)  & 74        & $75.8\pm3.8$ & $(18.4)$ & $49.4$ & $47.7\pm1.8$ & $(8.9)$ \\
Lead (Pb)     & 82        & $80.9\pm1.7$ & $(18.1)$ & $57.7$ & $65.1\pm2.2$ & $(23.4)$ \\
Uranium rods  & $\sim$65  & $59.4\pm2.5$ & $(20.5)$ & $\sim$55 & $63.2\pm2.0$ & $(16.5)$ \\
\end{tabular}
\end{center}
\caption{Reconstructed effective $Z$ and areal density values for the mock cargo materials for the 2000~s test ($\sim$0.33 mC integrated deuteron beam current on target
per cm of the scan).  Quoted uncertainty after the ``$\pm$'' represents the uncertainty on the average over the material region, while the
single pixel uncertainty is the standard deviation of the single pixel (1 cm$\times$1 detector) estimates over the region.}
\label{tab:short}
\end{table*}

\begin{figure*}[thb]
    \centering
    \includegraphics[width=1.5\columnwidth]{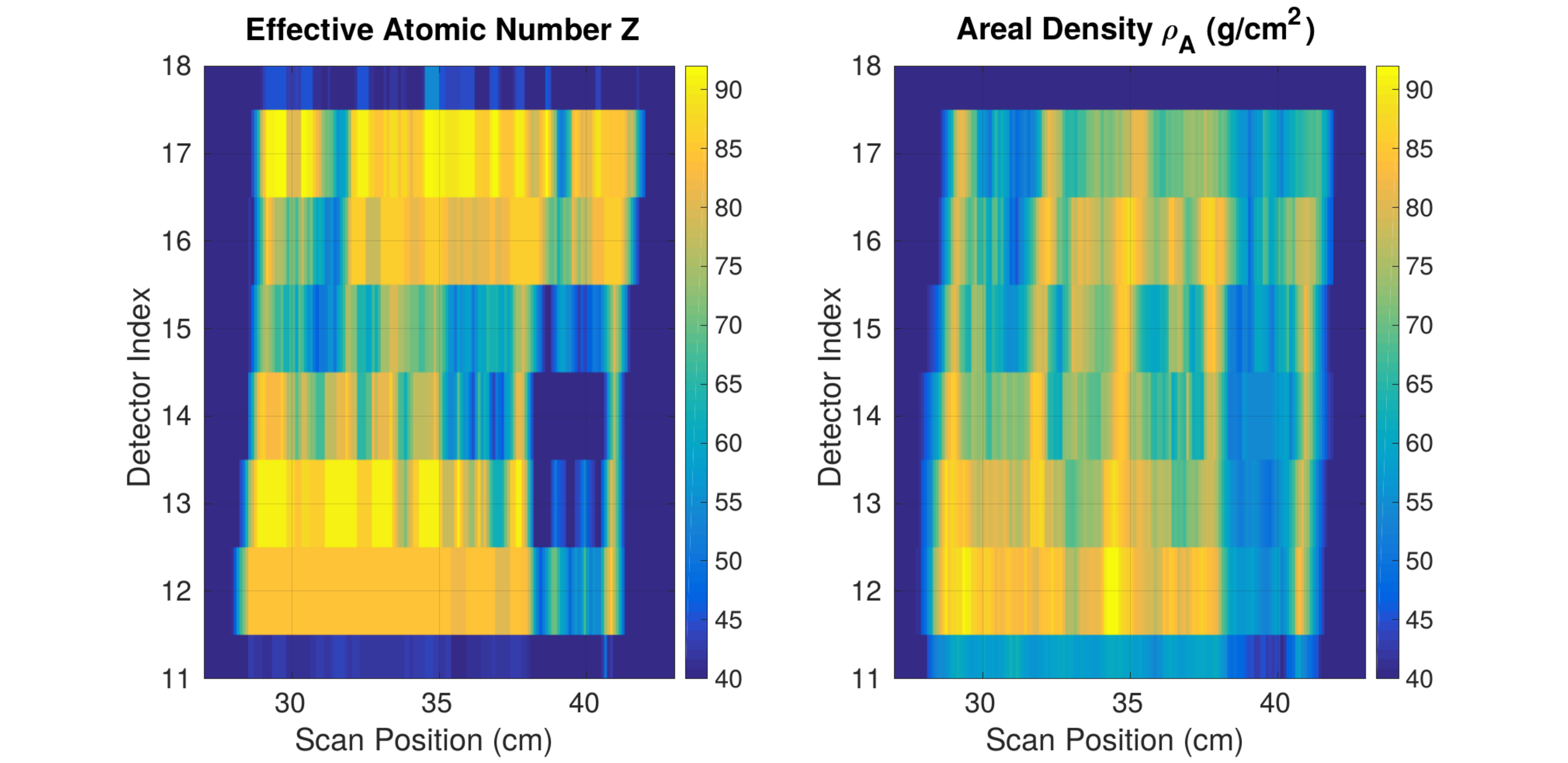}
    \caption{Images of the uranium rods in reconstructed effective $Z$ and areal density for the 7400~s test ($\sim$1.3 mC integrated deuteron beam current on target
per cm of the scan) with 1 mm wide pixels.}
    \label{fig:urods}
\end{figure*}

The results for the mixed material uranium rods merit further discussion.  Due to the fact that the rods consist of
aluminum and uranium, and additionally because they are not uniform in areal density as presented to the beam,
evaluation of the reconstruction of the material parameters for the rods is not as straightforward as for the pure
materials.  Appendix~\ref{app:eff} details rough estimates of the expected effective $Z$ and areal density for the
arrangement of the rods, up to the limited information available about the exact composition of the rods.
Due to the fact that the 1~cm pixel size in Figure \ref{fig:images} obscures the structures predicted by the results in Figure~\ref{fig:rodad},
it is useful to consider 1 mm pixels for the uranium rod sample despite the reduction in statistics.
Figure~\ref{fig:urods} shows the reconstructed $Z$ and $\rho_A$ for the rods with 1 mm pixels.  While the areal density
is slightly overestimated in the 1 mm pixels (due to low statistics), the images in Figure~\ref{fig:urods} clearly
show the structure of the rod arrangement (Figure~\ref{fig:rodarr}), and show that extra spacing between the rightmost
rods in combination with the uncertainty on the rod composition is likely responsible for the discrepancies between
the reconstructed $Z$ and $\rho_A$ values and the estimates from Appendix~\ref{app:eff}.  This mixed material example
demonstrates the limitations of 2D radiographic imaging to determine the material composition of cargo, but with sufficient
position resolution the presence of high-$Z$ material is still clearly evident.

\section{Conclusions and Future Work}

The results presented here establish the use of multiple monoenergetic gamma ray radiography (MMGR) to image
materials in both their effective atomic number and areal density.  Most notably, the technique distinguishes
pure materials even at high-$Z$ (e.g., separating Pb and W or Pb and U), a critical requirement of any
system designed to detect SNM and differentiate it from benign high-$Z$ materials, which could otherwise
result in false positive detections.  The specific information
transmitted by the monochromatic beam, combined with high resolution detectors to clearly % CHANGE
identify directly transmitted photons, provides the capability to identify materials while
minimizing the radiation dose delivered to the cargo.

The results for the natural uranium rods demonstrate the fundamental limitations of this technique, and
indeed radiography of any kind, as a method for detecting SNM.  The several mm of aluminum cladding
significantly reduces the effective $Z$ of the configuration, somewhat masking the presence of the uranium.
With sufficient position resolution, however, the presence of very high-$Z$ ($>$80) material can be flagged for this
configuration.  This suggests that monoenergetic gamma ray radiography may be paired with a secondary
technique (such as a system designed to detect induced photofission neutrons\cite{rose,pnpf-short}) to disambiguate
such situations.  Additionally, future work will explore the resolving power of radiography using multiple projections
for mixed material configurations.

A concern that has been raised regarding high energy gamma radiography techniques is the resulting
radiological activation of inspected materials.  Photons at 15.1 MeV have enough energy to induce $(\gamma,n)$ 
photodisintegration reactions in many elements, and may indirectly 
become a source of neutrons.  The capture of these secondary neutrons can 
transmute stable isotopes into metastable ones and induce long-lived 
radioactivity in the inspected materials.  Calculations show, however, that the 
exposure to the neutrons produced by the above reaction amounts to just one hour 
of exposure to cosmogenic neutrons from the natural background, and as such any 
contributions to induced radioactivity is negligible when compared naturally
occurring activation.  This calculation is detailed in Appendix~\ref{app:neutrons}.

The experimental setup used here would require significant modifications for deployment as a cargo scanning system,
several of which are the subject of ongoing work.  As discussed at the end of Section~\ref{sec:exp}, the scan times of several thousand
seconds used in this work would be reduced to $\lesssim$2 minutes by operating at mA-scale current.  The results presented here demonstrate
the ability of a radiography system to function in counting mode at such currents using the pile-up correction
technique detailed in Section~\ref{sec:pup}. In such a system, another technique would likely need to be devised to account
for the background subtraction conducted here using timing information.  Additionally, the use
of alternate nuclear reactions such as $^{12}$C$(p,p'\gamma)^{12}$C and $^{16}$O$(p,p'\gamma)^{16}$O, which produce
monoenergetic gamma rays between 4.4 and 8.9 MeV, would open a variety of options for different accelerators
and significantly reduce the neutrons that are present from other processes in a system using $^{11}$B(d,n$\gamma$)$^{12}$C.
Work is ongoing to establish the applicability of the techniques for precision material identification described
in this paper using the lower energy photons available from such reactions.

\section*{Acknowledgements}

This work is supported in part by the U.S. Department of Homeland Security Domestic Nuclear Detection Office under a
competitively awarded collaborative research ARI-LA Award, ECCS-1348328 and is part of a collaboration between the Massachusetts Institute of Technology,
Georgia Institute of Technology, University of Michigan, and Pennsylvania State University.
This support does not constitute an express or implied endorsement on the part of the United States Government.  The authors are grateful to Richard C. Lanza, who developed some of the initial ideas behind this work, for his support, encouragement, and valuable advice.
BSH gratefully acknowledges the support of the Stanton Foundation Nuclear Security Fellowship.
The authors wish to thank the MIT-Bates Research and Engineering Center staff for 
their invaluable contributions to the construction and operation of the experiment; in particular Peter Binns,
Hamid Moazeni, and Ernest Ihloff.  They thank Taylor Sims for his work on the experiment during
data taking.  Additionally, they thank Igor Jovanovic and Jayson Vavrek for valuable comments on the manuscript. 

\appendix

\section{Calculation of the Uranium Rod Effective $Z$ and Areal Density}
\label{app:eff}

\begin{figure}[thb]
    \centering
    \includegraphics[width=\columnwidth]{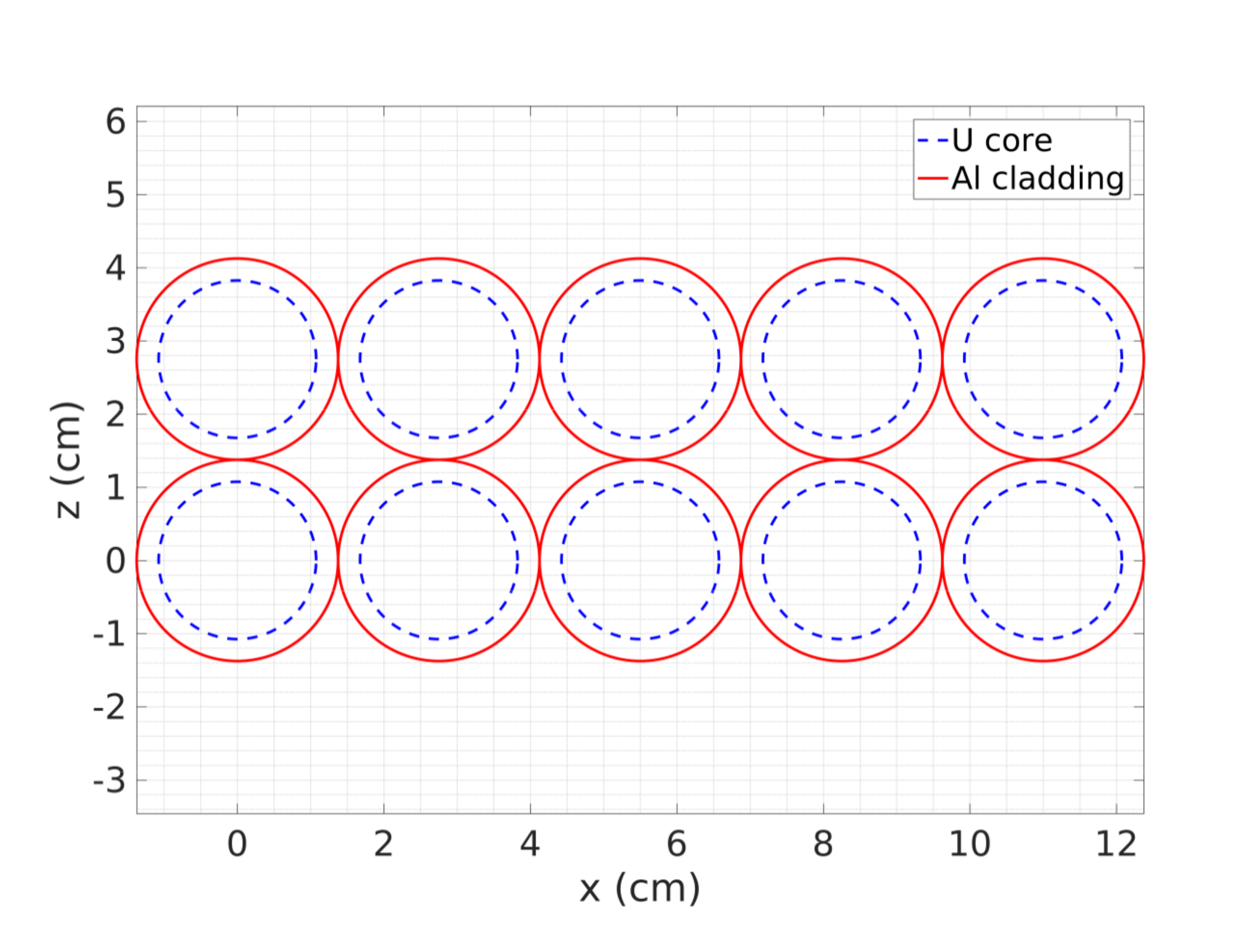}
    \caption{Arrangement of the 10 natural uranium rods used in the imaging test, viewed from above.
	     The beam was incident in the $+\mathbf{\hat{z}}$ direction.}
    \label{fig:rodarr}
\end{figure}

As described in Section~\ref{sec:mock}, the mock cargo objects for the imaging
demonstration results in this paper consisted of approximately uniform density
pure elemental materials except for the natural uranium rods.  These rods, used
on loan from the MIT Nuclear Reactor Laboratory, consisted of cylinders of natural
uranium metal clad with 0.3 cm of aluminum (for a total diameter of 2.75~cm).  For
the imaging test, 10 of these rods were arranged in two rows of five as shown in
Figure~\ref{fig:rodarr}.

% EDIT: Deciding to keep for now, can discuss

\begin{figure*}[thb]
    \centering
    \includegraphics[width=\columnwidth]{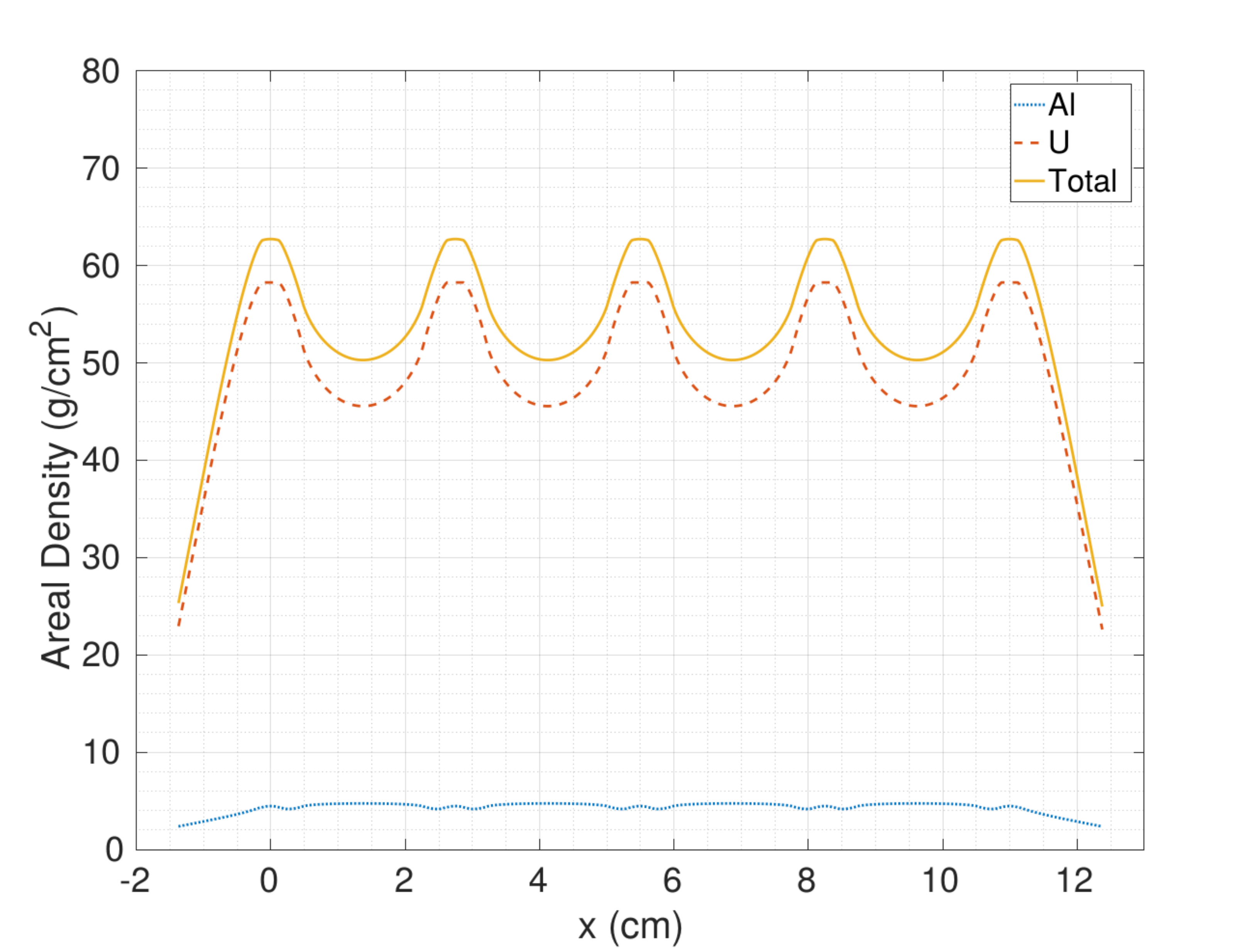}
    \includegraphics[width=\columnwidth]{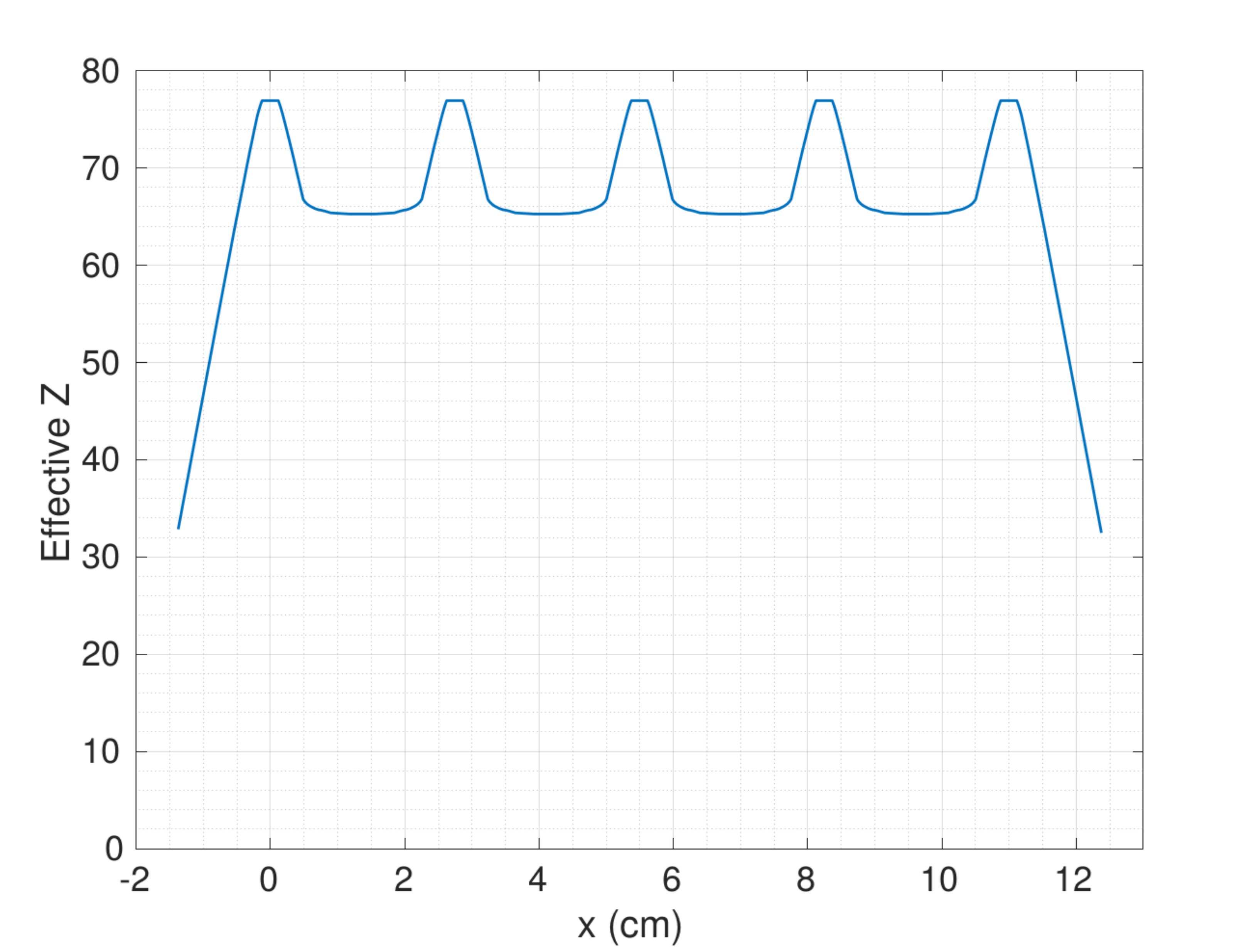}
    \caption{Expected reconstructed areal density (left) and effective $Z$ (right) of the uranium rod arrangement along the beam path, accounting
             for the loss of resolution due to the 2.38 cm width of the beam.}
    \label{fig:rodad}
\end{figure*}

Due to the fact that there was no means of verifying the internal composition
of the rods, the following calculations assume that the uranium metal inside
the cladding was of uniform density 19.3 g/cm$^3$.  Given this assumption, it
is possible to compute the areal density along the beam path across
the arrangement of the rods.  To compare to the imaging results, however,
the averaging effect of the finite collimator width must be considered.  At the
position of the mock cargo, the collimation produced a beam width of approximately
2.4 cm.  A sliding average over this width was applied to the exact calculation
of the areal density to produce the expected reconstructed areal density, shown
in Figure~\ref{fig:rodad}.

Computation of the expected effective $Z$ requires consideration of the nature
of the beam and physical processes that contribute to the attenuation of the
beam for the aluminum and uranium combinations across the arrangement.  To estimate
this value, the photon mass attenuation coefficients ($\frac{\mu}{\rho}$) for the uranium and aluminum
combinations across the configuration were estimated as functions of energy according to
\begin{equation}
 \label{eq:att}
 \frac{\mu}{\rho} = w_\text{U}\cdot \left(\tfrac{\mu}{\rho}\right)_\text{U} + (1-w_\text{U})\cdot \left(\tfrac{\mu}{\rho}\right)_\text{Al},
\end{equation}
where $w_\text{U}$ is the uranium mass fraction at a given position, as computed above, and the attenuation coefficient values for the individual
elements provided by the NIST x-ray mass attenuation coefficient database \cite{xray}.  For each position in the configuration,
the pure element with attenuation coefficient values at 4.4 MeV and 15.1 MeV best matching the rod attenuation coefficients
as computed by Equation~\ref{eq:att} was used as an estimate of the effective value of $Z$ at that position.  The results
of this estimate, averaged over the beam illumination width, are also shown in Figure~\ref{fig:rodad}.

% \begin{figure}[thb]
%     \centering
%     \includegraphics[width=\columnwidth]{figures/rodz.pdf}
%     \caption{Expected reconstructed effective $Z$ of the uranium rod arrangement along the beam path, accounting
%              for the loss of resolution due to the 2.38 cm width of the beam.}
%     \label{fig:rodz}
% \end{figure}

\section{Calculation of the Pile-Up Correction Factor}
\label{app:pcor}

In this appendix, the scaling factor that must be applied to a pile-up
rejected (``clean'') spectrum with counts $N_C$ is computed for both paralyzable and
non-paralyzable deadtime scenarios.  Since the true rate of individual events $r$ is unknown, and
the resolving time of the detector $\tau$ may also be unknown, it is most useful
to express the correction as a function of the fraction of total counts captured
in the pile-up rejected spectrum $f$ (i.e., the ``clean'' fraction of events).

To compute the correction, let $N_\text{rec}$ be the total number of recorded counts (including pile-up rejected counts), $N_\text{c}$ be the number of counts recorded
cleanly without pile-up, and $N_\text{pu}$ be the number of counts in the pile-up
portion of the spectrum (or the number of rejected counts). That is, $N_\text{rec}=N_\text{c}+N_\text{pu}$).
The number of true counts $N$ may then be expressed as
\begin{equation}
 N = N_\text{c}+N_\text{pu}\cdot\overline{n},
 \label{eq:bcor}
\end{equation}
where $\overline{n}$ is the mean number of true counts per pile-up count.  Since
the clean spectrum represents the correct shape of the true spectrum, it is useful
to express Equation~\ref{eq:bcor} as a scaling of the number of clean counts
as the fraction $f=\frac{N_\text{c}}{N_\text{rec}}$ of the total recorded counts that they
represent:
\begin{equation}
 N = N_\text{c}\left(1 + \frac{N_\text{rec}-N_\text{c}}{N_\text{c}}\cdot\overline{n} \right) = N_\text{c}\left(1 + \frac{1-f}{f}\cdot\overline{n} \right).
 \label{eq:cor}
\end{equation}
The true spectrum may then be recovered by scaling the extracted clean spectrum,
given the mean number of true counts per pile-up count $\overline{n}$.  Let $P(j)$
be the probability that given a single true event, $j$ additional events are recorded
as piled-up with the original event (i.e., $j=0$ represents a cleanly recorded event). Then
the mean number of counts per piled-up count is:
\begin{equation}
 \overline{n} = \frac{1}{\sum_{j=1}^{\infty}P(j)} \sum_{j=1}^{\infty}\left(j+1\right)P(j) = \frac{1}{1-f} \sum_{j=1}^{\infty}\left(j+1\right)P(j),
 \label{eq:nbar}
\end{equation}
since the term in the denominator is simply the probability that the original count
accumulates at least one additional count when recorded.

The following sections compute $\overline{n}$ as a function of $f$ for the cases of non-paralyzable
and paralyzable detector and data acquisition systems, re-expressing the standard
results for pile-up rates \cite{knoll} to eliminate need for knowledge of the true detector rates
and resolving times. 

\subsection{Non-Paralyzable Case}

For a non-paralyzable detector and data acquisition system exposed to a
true rate of incident events $r$ with an integration time of $\tau$ per
recorded event, the probability $P(j)$ of $j$ true events following a given event
trigger within the same trigger window
is governed by the Poisson distribution \cite{knoll}:
\begin{equation}
 P(j) = \frac{\left(r\tau\right)^j}{j!}e^{-r\tau}.
 \label{eq:pois}
\end{equation}
Combining Equations~\ref{eq:nbar} and~\ref{eq:pois} for the non-paralyzable case
yields
\begin{equation}
 \overline{n}_\text{NP} = \frac{e^{-r\tau}}{1-f} \sum_{j=1}^{\infty}\left(j+1\right)\frac{\left(r\tau\right)^j}{j!}.
\end{equation}
Noting that $\exp(x) = \sum_{j=0}^\infty x^j/j!$, the sums reduce to
\begin{equation}
 \overline{n}_\text{NP} = \frac{1}{1-f}\left(1-e^{-r\tau}+r\tau\right).
\end{equation}
As noted, it is most useful to recast
the expression as a function of the fraction of the total counts in the pile-up rejected spectrum $f= P(0) = e^{-r\tau}$, to yield
\begin{equation}
 \overline{n}_\text{NP} = 1-\frac{\ln f}{1-f}.
 \label{eq:nnp}
\end{equation}
Note that $\overline{n}_\text{NP}$ monotonically increases with the piled-up fraction
($1-f$) and $\lim_{f\to1}\overline{n}_\text{NP}=2$, i.e., piled-up events contain a negligible fraction
of multiple pile-up ($j>1$) as the pile-up rate goes to zero.

\subsection{Extending Case}

In the paralyzable case, the arrival of each new true event
resets the integration time $\tau$ per event, leading to a greater
probability of accumulating $j>1$ piled-up true events along with the event
that initialized the trigger.  The probability of
$j$ events piling-up with a given single event which initializes a trigger for the
paralyzable system case is\cite{knoll}
\begin{equation}
 P(j) = e^{-r\tau}\left(1-e^{-r\tau}\right)^j.
 \label{eq:ext}
\end{equation}
Note that, as in the non-paralyzable case, $f=P(0)=e^{-r\tau}$.  Combining Equations
\ref{eq:nbar} and~\ref{eq:ext} yields
\begin{equation}
 \overline{n}_\text{P}  = \frac{f}{1-f} \sum_{j=1}^{\infty}\left(j+1\right)\left(1-f\right)^j .
\end{equation}
Since $0\leq f\leq1$, the sum is an arithmetico-geometric series:
\begin{equation}
 \overline{n}_\text{P} = 1+\frac{1}{f}.
 \label{eq:np}
\end{equation}
This quantity also approaches $\overline{n}=2$ for $f\to1$ like Equation~\ref{eq:nnp}, but increases
much more rapidly as $f$ goes to zero, as expected for the extended integration
window.

\section{Neutrons from the $(\gamma,n)$ reactions}
\label{app:neutrons}

The 15.1 MeV photons in the presented system have energies well above the photodisintegration thresholds 
for a number of common materials, and thus can produce neutrons via the 
$X(\gamma,n)Y$ reaction.  The resulting radiological activation of inspected
materials resulting from transmutation of nuclei into unstable isotopes as
these neutrons are captured has been a commonly
cited concern regarding the use of high energy gamma rays for cargo inspection\cite{photoneutron}.
To understand the scale of this effect, it is useful to estimate the production rate
for these neutrons and compare it to the rate of exposure to cosmogenic 
neutrons.

As an example, consider the case of a plate of steel 2.4 m tall,  undergoing a
scan at the speed of 40 cm/s with 15.1 MeV photons.  The rate of neutron
production in the steel is
\begin{equation}
N = \frac{\sigma  N_A}{A} \phi_0 \frac{1}{\mu } [1-e^{-\mu \rho D}],
\end{equation}
where $N$ is the neutron production rate, $\sigma$ is the $(\gamma,n)$ total 
cross section, $N_A = 6.022\times10^{23}$ is Avogadro's number, $A$ is the atomic 
weight of the target material,  $\phi_0$ is the incident photon rate, $\mu$ is the mass attenuation 
cross section, $\rho$ is the density, and $D$ is the thickness of the material.

For a realistic cargo scanning system using the $^{11}$B(d,n$\gamma$)$^{12}$C reaction with 1~mA of deuteron
current, $\phi_0 \approx 2\times10^{10}\:\text{mA}^{-1}\text{s}^{-1}$(\cite{buck}), assuming a collimation
width of 3~mm at a distance 1~m from the target.
Taking $D \gg 1/\mu \rho$, i.e. a thick slab,  $\sigma=40 $ mb(\cite{nndcurl}), and $\mu=0.031\:\text{cm}^2/\text{g}$(\cite{xcom}), 
the neutron production rate would be $N=4.2\times 10^5\ $s$^{-1}$.
For a comparison, the cosmogenic neutron rate is estimated by Gordon {\it et 
al.}\cite{gordon2004measurement} is approximately 0.0134~cm$^{-2}$s$^{-1}$, which
includes contributions from thermal neutrons in addition to the fast
neutrons of interest for this comparison.  
Integrating over the volume of the plate described above, the cosmogenic neutron
exposure rate is $N_{cosm.} \approx 130$~s$^{-1}$.
The neutron production by $(\gamma,n)$ reactions in the steel plate is approximately 3300 times higher than the exposure to 
cosmogenic neutrons, which corresponds to only $\sim$1 hour of effective additional exposure to natural 
neutron background.  As such, the exposure to the high energy gamma rays in a single scan has a negligible effect
on the additional buildup of transmuted elements and activation of the material.

%%%%%%%%%%%%%%%%%%%%%%%%%%%%%%%%%%%%%%%%%%%%%%%%%%%%%%%%%%%%%%%%%%%%%%%%%%%%%%%%
% \tableofcontents

%%%%%%%%%%%%%%%%%%%%%%%%%%%%%%%%%%%%%%%%%%%%%%%%%%%%%%%%%%%%%%%%%%%%%%%%%%%%%%%%
%
% Practical notes:
%
% Submit to physics and society on arXiv
%
%

%\bibliographystyle{ieeetr}
%\nocite{*}
\bibliography{references}

\end{document}